\documentclass{emulateapj}

\usepackage{float}
\usepackage{amsmath}
\usepackage{epsfig,floatflt}
\usepackage{subfigure}

    \newcommand{\qed}{\nobreak \ifvmode \relax \else
          \ifdim\lastskip<1.5em \hskip-\lastskip
          \hskip1.5em plus0em minus0.5em \fi \nobreak
          \vrule height0.75em width0.5em depth0.25em\fi}

\newcommand{\commander}{{\tt Commander}}

\begin{document}

\title{Bayesian component separation and CMB estimation for the 5-year WMAP temperature data}

\author{C. Dickinson\altaffilmark{1,2,3}, H. K. Eriksen\altaffilmark{4,5}, A. J. Banday\altaffilmark{6}, J. B. Jewell\altaffilmark{7}, K. M. G\'{o}rski\altaffilmark{7,8,9}, G. Huey\altaffilmark{7}, C. R. Lawrence\altaffilmark{7,8} \\
I. J. O'Dwyer\altaffilmark{7}, B. D. Wandelt\altaffilmark{10,11,8}}

\altaffiltext{1}{email: Clive.Dickinson@manchester.ac.uk}
\altaffiltext{2}{Infrared Processing \& Analysis Center, California Institute of Technology, M/S 220-6, 1200 E. California Blvd., Pasadena, CA 91125.}
\altaffiltext{3}{Current address: Jodrell Bank Centre for Astrophysics, Department of Physics \& Astronomy, Alan Turing Building, University of Manchester, Oxford Rd., Manchester, M13 9PL, U.K.}
\altaffiltext{4}{Institute of Theoretical Astrophysics, University of
Oslo, P.O.\ Box 1029 Blindern, N-0315 Oslo, Norway}
\altaffiltext{5}{Centre of
Mathematics for Applications, University of Oslo, P.O.\ Box 1053
Blindern, N-0316 Oslo}
\altaffiltext{6}{Max-Planck-Institut f\"ur Astrophysik,
Karl-Schwarzschild-Str.\ 1, Postfach 1317, D-85741 Garching bei
M\"unchen, Germany}
\altaffiltext{7}{Jet Propulsion Laboratory, 4800 Oak
  Grove Drive, Pasadena CA 91109} 
\altaffiltext{8}{California Institute of Technology, Pasadena, CA
  91125} 
\altaffiltext{9}{Warsaw University Observatory, Aleje Ujazdowskie 4, 00-478 Warszawa, Poland}
\altaffiltext{10}{Department of Physics, University of Illinois, Urbana, IL 61801.}
\altaffiltext{11}{Astronomy Department, University of Illinois, Urbana, IL 61801-3080.}

\date{Received - / Accepted -}

\begin{abstract}
A well-tested and validated Gibbs sampling code, that performs component separation and CMB power spectrum estimation, was applied to the {\it WMAP} 5-yr data. Using a simple model consisting of CMB, noise, monopoles and dipoles, a ``per pixel'' low-frequency power-law (fitting for both amplitude and spectral index), and a thermal dust template with fixed spectral index, we found that the low-$\ell$ ($\ell < 50$) CMB power spectrum is in good agreement with the published {\it WMAP}5 results. Residual monopoles and dipoles were found to be small ($\lesssim 3~\mu$K) or negligible in the 5-yr data. We comprehensively tested the assumptions that were made about the foregrounds (e.g. dust spectral index, power-law spectral index prior, templates), and found that the CMB power spectrum was insensitive to these choices. We confirm the asymmetry of power between the north and south ecliptic hemispheres, which appears to be robust against foreground modeling. The map of low frequency spectral indices indicates a steeper spectrum on average ($\beta=-2.97\pm0.21$) relative to those found at low ($\sim$GHz) frequencies.
\end{abstract}

\keywords{cosmic microwave background --- cosmology: observations --- radio continuum: ISM}

\maketitle


\section{Introduction}

Accurate separation of diffuse Galactic foreground emissions is one of
the major challenges of cosmic microwave background (CMB) data
analysis. Foreground emission must be subtracted accurately and errors
should be correctly propagated through the data analysis pipeline to extract
precise cosmological information. Furthermore, the use of Galactic
masks reduces the sky area that can be exploited for cosmological
analyses thus increasing the sample variance at large angular
scales. At angular scales $> 1^{\circ}$, the dominant foregrounds are
from synchrotron, free-free and vibrational dust emission from the
Galaxy. In addition, there is evidence for an additional ``anomalous''
foreground at low frequencies ($\sim10-60$~GHz), which is still not well understood \citep{davies:2006,dickinson:2006,dickinson:2007,dickinson:2009,bonaldi:2007,hildebrandt:2007,dobler:2008a,gold:2009}.

Based on earlier pioneering work by \cite{jewell:2004} and \cite{wandelt:2004}, \cite{eriksen:2008a} describe a Gibbs sampling algorithm that can
perform component separation and power spectrum estimate {\it at the same
time}, by sampling from the joint posterior probability distribution for
all fitted parameters i.e. it is based on Bayesian statistics.  This
method has the advantage of full propagation of error information
through to the CMB power spectrum (errors for each $C_\ell$ estimate)
but also allows considerable flexibility in fitting of foreground
model parameters. These can include fixed index templates, free
templates at each frequency, monopoles/dipoles, and any spectral
function (e.g. power-law at each pixel). A detailed description of the
algorithm and the implementation of the parallelized Fortran90 code
(henceforth, \commander) is given by \cite{eriksen:2008a}. The code also takes into accounts finite bandpass responses at each frequency.

In a previous work \citep{eriksen:2008b}, \commander~was used analyze the
3-year {\it WMAP} data \citep{hinshaw:2007}, at a resolution of $3^{\circ}$
(N$_{\rm side}=64$). The analysis confirmed the accuracy of the CMB
power spectrum up to $\ell=30$ and provided estimates of the
foreground components including a map of the low frequency foreground
amplitude and spectral index at each pixel. Due to the limited number
of frequency channels and frequency range of the {\it WMAP} data, a number
of assumptions had to be made such as using a thermal dust template
with fixed spectral index and a prior on the foreground spectral
index. Furthermore, they discovered the existence of
significant monopoles (offsets) and dipoles in the 3-year {\it WMAP}
data. Although this did not significantly affect the cosmological
interpretation (but did affect the details of the foregrounds) it was
later confirmed to be an error in the processing of the 3-year data \citep{hinshaw:2009}.

In this paper we analyze the new {\it WMAP} 5-year data using the \commander~code and confirm the CMB results from the {\it WMAP} team. \S\ref{sec:data_foregrounds} outlines the data and pre-processing along with a brief summary of diffuse foreground components and the templates and models that are used to fit them. The basic {\it WMAP}5 analysis is presented in \S\ref{sec:wmap5_basic}. In \S\ref{sec:varying_models} we investigate the sensitivity of the results on a variety of different foreground models. We show that the results are largely insensitive to the assumptions made. A comparison of low-$\ell$ CMB and foreground spectral index maps is made in \S\ref{sec:comparison}. Conclusions are given in \S\ref{sec:conclusions}.


\section{DATA}
\label{sec:data_foregrounds}

\subsection{{\it WMAP} data and preprocessing steps}

The main input data are the 5-yr {\it WMAP} temperature maps provided on Lambda\footnote{http://lambda.gsfc.nasa.gov/}. We perform the same pre-processing steps taken by \cite{eriksen:2008b}, to allow the Gibbs sampling to be performed relatively efficiently. The $1^{\circ}$ smoothed maps frequency maps were further smoothed to a common resolution of $3^{\circ}$ and downgraded to a HEALPix\footnote{http://healpix.jpl.nasa.gov} resolution of N$_{\rm side}=64$, corresponding to a pixel size of $55\arcmin$. The single frequency band maps are at nominal frequencies of 23, 33, 41, 61 and 94~GHz. We assume a top-hat bandpass response function with widths given by the radiometer specifications in \cite{jarosik:2003}. An additional $2~\mu$K white noise is added to the maps to regularize the noise covariance matrix at high $\ell$-values and thus reduce the number of conjugate gradient descent (CGD) iterations when solving the map-making equation. The additional noise has a negligible effect for $\ell \lesssim 50$. 

Note that the amount of smoothing and additional noise is somewhat arbitrary, and was a choice that was made to balance the accuracy of the results and the amount of CPU time. A typical analysis takes  $\sim 1$~minute per Gibbs iteration on a modern 3~GHz CPU. Thus for a single run, with 10000 Gibbs samples distributed over 4 chains ($4\times5=20$ CPUs in total), takes a total of $\sim 40$ hours. We note that \cite{jewell:2008} have recently developed a Markov Chain Monte Carlo (MCMC) algorithm that effectively solves the problems associated with direct Gibbs sampling in the low signal-to-noise regime. 

\subsection{Foreground components and templates}

Diffuse foreground emission consists of at least 3 well-known components: synchrotron, free-free and thermal (vibrational) dust. In additional to this, there is significant evidence for an anomalous component, possibly due to spinning dust. We briefly review the components and simple models that are used when fitting to real data.

Synchrotron radiation is emitted by relativistic cosmic ray electrons spiralling in the Galactic magnetic field. If the energy spectrum of the cosmic rays is a power-law, $dN/dE=N_0 E^{-p}$, then the intensity is given by $I(\nu)=LN_0B_{\rm eff}^{(p+1)/2}\nu^{-(p-1)/2}a(p)$, where $a(\nu)$ is a weak function of frequency, $L$ is the length along the line-of-sigh of the emitting volume, $B_{\rm eff}$ is the effective magnetic field strength and $\nu$ is the frequency. A power-law, $T(\nu)=A\nu^{\beta_{\rm s}}$, is therefore a reasonable approximation for synchrotron radiation with $\beta_{\rm s}\sim -2.7$ at radio wavelengths. At higher frequencies, a single source of electrons will lose energy faster (spectral aging) resulting in a steepening of the spectral index, by $\sim 0.5$. However, multiple synchrotron components along the same line-of-sight will tend to result in a flattening of the spectral index at higher frequencies. At {\it WMAP} frequencies, the values of $\beta_{\rm s} \approx -3.0 \pm 0.3$ have been reported. The steep spectral index makes low frequency radio maps ideal templates for synchrotron radiation. The most widely used is the all-sky 408~MHz map of \cite{haslam:1982}.

Free-free (bremsstrahlung) emission is emitted by free electrons in warm ($T_e\approx 10^4$~K) gas. At radio frequencies,the radio spectrum has the form $T_{\rm ff} \propto T_e^{-1.35}\nu^{-2.1}EM$ where $EM$ is the Emission Measure. The spectrum therefore closely follows a power-law with a spectral index of $\beta_{\rm ff}=-2.1$. A more accurate calculation, taking into account the Gaunt factor, gives a slightly steeper index at {\it WMAP} frequencies, with $\beta_{\rm ff}\approx -2.15$ and some weak dependence on $T_e$ and other environmental conditions \citep{dickinson:2003}. However, without a large lever arm in frequency, the separation of synchrotron and free-free emission can be difficult. An alternative method is to use a spatial template of the free-free. Recombination lines (e.g. H$\alpha$) are also proportional to $EM$ and therefore provide a reliable tracer of free-free emission, if $T_e$ is known; at high latitudes, $T_e \approx 8000$~K. The major limitation when using optical lines is the visual dust extinction. Fortunately, at high Galactic latitudes, the extinction by dust is small. Full-sky maps are now available including corrections for dust extinction \citep{finkbeiner:2003,dickinson:2003}.

Thermal (vibrational) dust emission arises from the heating of dust grains by the UV interstellar radiation field. At an equilibrium temperature of $T_{\rm d} \approx 18$~K, it emits mostly in the far infra-red, peaking at $\lambda \approx 100~\mu$m. For CMB experiments, it is only a major foreground at $\nu \gtrsim 100$~GHz. A common approximation to the thermal dust spectrum is a modified black-body: $T(\nu) \propto \nu^{\epsilon}B(\nu,T_{\rm d})$ where $\epsilon$ is the emissivity index, with typical values $\sim 2$ (Table~\ref{tab:models}). At {\it WMAP} frequencies, we see the Rayleigh-Jeans tail of the thermal dust, and hence the spectrum can be approximated by a power-law, based on the emissivity index, $\epsilon$. However, thermal dust is relatively weak at {\it WMAP} frequencies, therefore spatial templates are often used as a tracer instead. The most widely used template is the dust prediction of \cite{finkbeiner:1999}, relative to 94~GHz. Based on IRAS/DIRBE data, it represents a global fit to the all-sky data, with 2 dust components with different temperatures and emissivities. 

Anomalous emission is still not well understood. It appears to be an excess of emission at $\sim 10-60$~GHz and is tightly correlated to FIR dust emission. The best explanation is in terms of electro-dipole radiation from rapidly spinning small dust grains. Theoretical models predict a peaked (convex) spectrum with a peak frequency in the range $\sim 20-40$~GHz \citep{draine:1998a,draine:1998b,ali-hamoud:2008}. The anomalous emission is difficult to separate spectrally in the {\it WMAP} channels since it is falling off quickly in the lower {\it WMAP} bands, with a spectral similar to synchrotron. As with free-free, a separation based on spatial morphology has been the most successful way to separate the anomalous emission with the other components, due to the remarkably tight correlation with FIR templates \citep{davies:2006}.
 
In this paper, we use standard foreground templates to trace diffuse foreground components (also downloaded from the Lambda website). For the synchrotron emission, we use the destriped and source-removed removed NCSA version of the \cite{haslam:1982} 408~MHz all-sky map. For free-free emission, we use the all-sky dust-corrected H$\alpha$ map produced by \cite{finkbeiner:2003}. We check the consistency with the similar template by \cite{dickinson:2003}. For thermal dust, we use the extrapolation of model 8 of \cite{finkbeiner:1999} (hereafter FDS99) to 94~GHz. We check the consistency with the original $100~\mu$m intensity map, and also the dust reddening $E(B-V)$ map. All maps were smoothed to a common $3^{\circ}$ resolution and downgraded to $N_{\rm side}=64$.

\begin{table*}
\centering
\caption{Signal components and models. Spectra are given in terms of a Rayleigh-Jeans temperature, $T_{\rm RJ}$.}
\label{tab:models}
  \begin{tabular}{ll}
    \hline
Component      &Models   \\
\hline

CMB            &1. Gaussian random field with known frequency spectrum. \\
               &2. Uncorrelated pixels, CMB spectrum: $T_{\rm RJ} = A \times (e^{h \nu / kT}-1)^2/(({h \nu / kT})^2 e^{h \nu / kT})$  \\ \hline

Synchrotron    &1. Spatial template: Haslam et al. 408~MHz all-sky map  \\
               &2. Spatial template: $K-Ka$ {\it WMAP} difference map \\
               &3. Uncorrelated pixels, power-law: $T_{\rm RJ} = A \times (\nu/\nu_{0})^{\beta_{\rm s}}$ \\ 
               &4. Uncorrelated pixels, power-law with running: $T_{\rm RJ} = A \times (\nu/\nu_{0})^{\beta_{\rm s} + C{\rm log}(\nu/\nu_{0})}$ \\  \hline

Free-free      &1. Spatial template: H$\alpha$ maps \citep{finkbeiner:2003,dickinson:2003} \\
               &2. Uncorrelated pixels, fixed index power-law: $T_{\rm RJ} = A \times (\nu/\nu_{0})^{-2.15}$ \\ \hline

Anomalous dust &1. Spatial template: Dust map extrapolated to 94~GHz \citep{finkbeiner:1999} \\
               &2. Spatial template: $K-Ka$ {\it WMAP} difference map \\
               &3. Uncorrelated pixels, power-law:  $T_{\rm RJ} = A \times (\nu/\nu_{0})^{\beta}$ \\ \hline

Thermal dust   &1. Spatial template: Dust map extrapolated to 94~GHz \citep{finkbeiner:1999} \\
               &2. Uncorrelated pixels, power-law:  $T_{\rm RJ} = A \times (\nu/\nu_{0})^{\beta_{\rm d}}$ \\ 
               &3. Uncorrelated pixels, modified black-body $T_{\rm RJ}=A \times \nu^{\epsilon} B(\nu,T_{\rm d})$ \\ 
\hline      
 \end{tabular}
\end{table*}


\section{Basic {\it WMAP}5 analysis}
\label{sec:wmap5_basic}

\subsection{Signal model and priors}

To begin with, we apply the signal model as used by \cite{eriksen:2008b} to the {\it WMAP} 5-yr data. For this case, the signal model being fitted to the data, is of the form:-
\begin{eqnarray*}
T_\nu(p) = s(p) + m_\nu^0 + \sum_{i=1}^3 m_\nu^i[\hat{{\bf e}}_{i} \cdot \hat{{\bf n}}(p)] \\ 
+ b\left[t(p)a(\nu)\left(\nu/\nu_0^{\rm dust}\right)^{\beta_{\rm d}}\right] + f(p)a(\nu)\left(\nu/\nu_0\right)^\beta(p)
\end{eqnarray*}

The thermodynamic temperature $T_\nu(p)$, in each pixel $p$, consists of terms from the CMB (assumed to be Gaussian and defined by a power spectrum up to $\ell=150$), monopoles and dipoles at each frequency, a single dust template amplitude, $b$, based on the FDS99 model 8 prediction $t(p)$ and with a fixed index of $\beta_{\rm d}=+1.7$, and a single power-law (amplitude and spectral index) for each pixel in the R-J convention; $a(\nu)$ is the conversion from thermodynamic to R-J temperature units. The power-law accounts for all the low frequency foreground emission. We adopt a weak uniform prior on the power-law spectral index of $-4.0 < \beta < -1.0$ and a more constraining Gaussian prior with mean $\beta=-3.0$ and standard deviation $\Delta \beta=0.3$. These are multiplied by the appropriate Jeffreys prior to account for the different volume of likelihood space for different spectral indices \citep{eriksen:2008a}. The mask is a downgraded version of the standard Kp2 mask (not including the majority of extragalactic sources) where any masked sub-pixels result in a masked pixel at $N_{\rm side}=64$, resulting in $14.4\%$ of pixels being masked.

\subsection{CMB results}

\begin{figure*}[]
\begin{center}
\mbox{\epsfig{figure=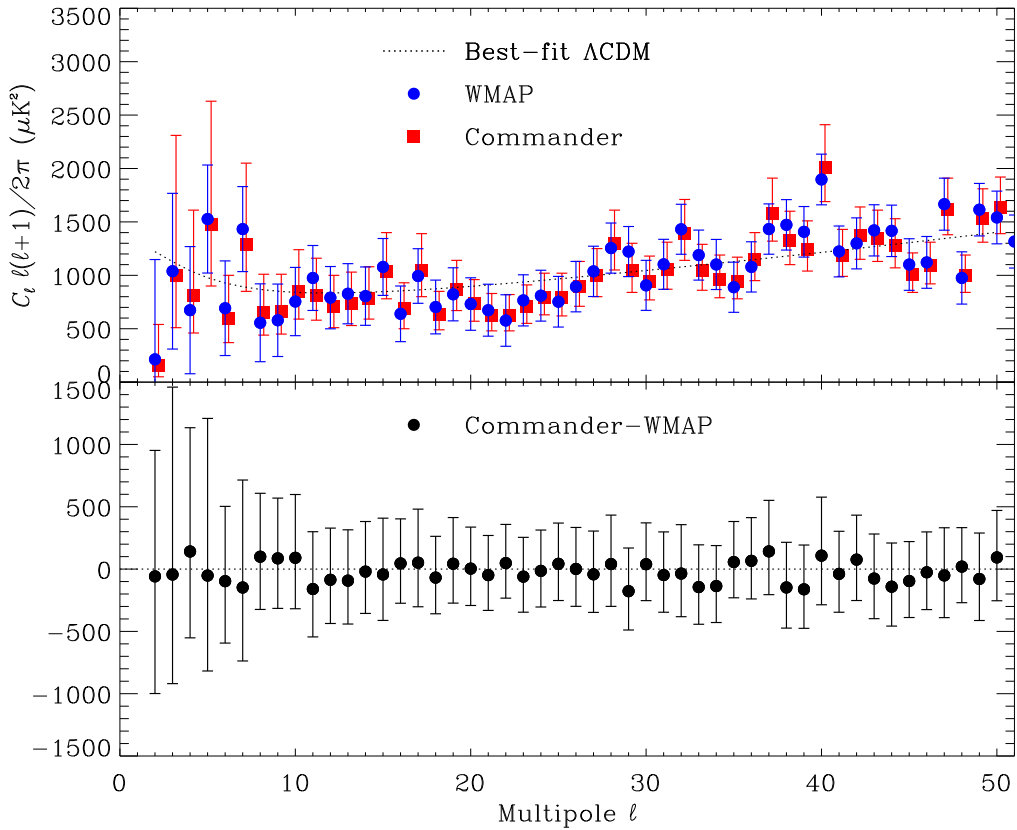,width=0.8\linewidth,clip=,angle=0}}
\caption{{\it Top}: CMB temperature power spectrum obtained by \commander~({\it red filled squares}) and the {\it WMAP} team ({\it black filled circles}) based on the {\it WMAP} 5-yr data. The best-fit $\Lambda$CDM model spectrum of \cite{nolta:2009} is shown as a dotted line. The error bars show asymmetric $68\%$ confidence limits for \commander, and Gaussian $1\sigma$ uncertainties for the {\it {\it WMAP}} spectrum. {\it Bottom}: Difference between the {\it WMAP} team and \commander~power spectra. }
\label{fig:cls_wmap5}
\end{center}
\end{figure*}

The CMB temperature power spectrum from \commander~and the {\it WMAP} team is shown in Fig.~\ref{fig:cls_wmap5}. The maximum likelihood value and $68\%$ confidence regions were calculated using the ``Gaussianized Blackwell-Rao'' estimator of \cite{rudjord:2008}. The power spectrum is in very good agreement with those from {\it WMAP} 3-yr data and those from the {\it WMAP} team \citep{nolta:2009}, on a $\ell$-by-$\ell$ basis. The error bars are not directly comparable since we calculate the $68\%$ confidence limit for each $C_\ell$ while we plot just the Gaussian symmetric error bars from the {\it WMAP} team. 

\begin{figure*}[]
\begin{center}
\mbox{\epsfig{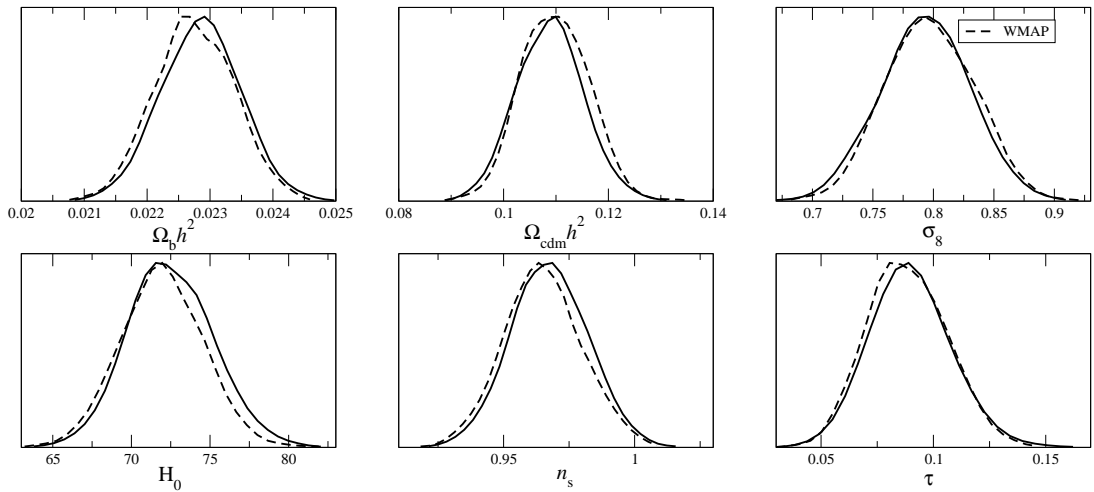}}
\caption{Comparison of marginal probability distributions from {\it WMAP} 5-yr data of \commander~(solid line) and the {\it WMAP} team (dashed line) for a six-parameter cosmological model.}
\label{fig:wmap5_params}
\end{center}
\end{figure*}

To determine the impact on the cosmological parameters we ran the derived {\it WMAP}5 power spectrum through {\sc cosmomc}\footnote{http://cosmologist.info/cosmomc/} \citep{lewis:2002} for the standard 6-parameter $\Lambda$CDM model \citep{dunkley:2009}. In a similar way to the {\it WMAP} team, and also \cite{chu:2005}, we use the Blackwell-Rao estimates for $\ell<32$, and fix the $C_{\ell}$'s for $\ell=33-50$ during the run, corresponding to a constant in the log-likelihood. The \commander~and {\it WMAP} likelihoods for the six-parameter set are compared in Fig.~\ref{fig:wmap5_params}. The distributions are almost identical. The \commander~marginal means and standard deviations for these parameters are $\Omega_{\rm b}h^2=0.02285\pm0.00063$, $\Omega_{\rm cdm}h^2=0.1089\pm 0.0064$, $\sigma_{8}=0.793\pm0.037$, $H_0=72.5\pm 2.7$, $n_{s}=0.9675\pm 0.0146$ and $\tau=0.0898\pm 0.0177$. These are in excellent agreement with the {\it WMAP} 5-yr likelihood values: $\Omega_{\rm b}h^2=0.02278\pm0.00062$, $\Omega_{\rm cdm}h^2=0.1093\pm 0.0067$, $\sigma_{8}=0.793\pm0.038$, $H_0=72.2\pm 2.7$, $n_{s}=0.9645\pm 0.0147$ and $\tau=0.0891\pm 0.0175$.  

\subsection{Foreground results}

\begin{figure*}[t]
\mbox{\epsfig{figure=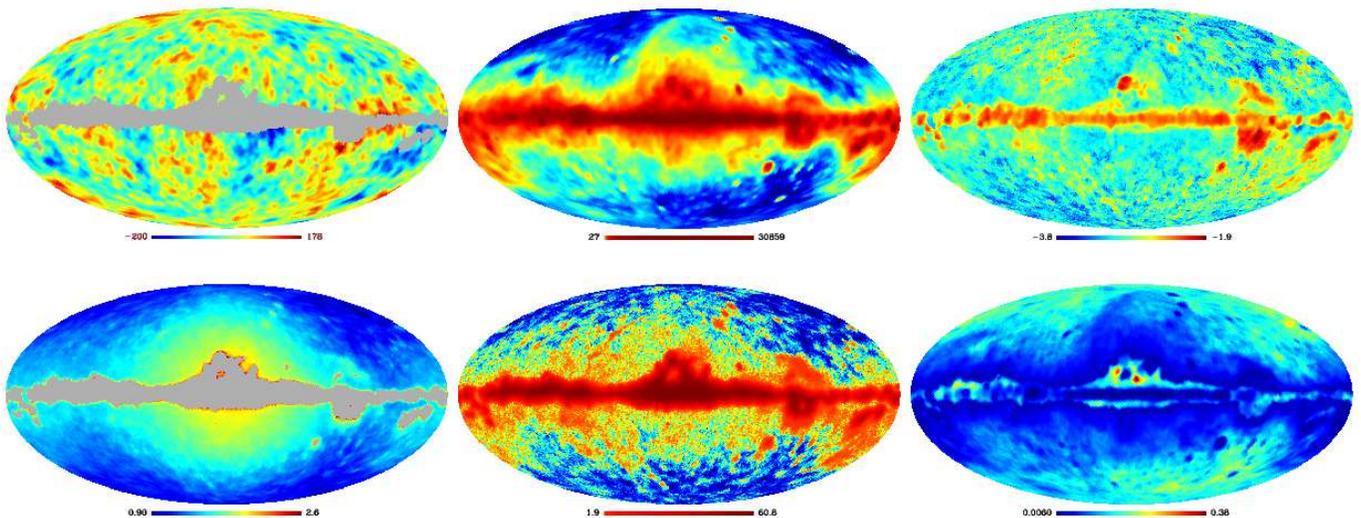,width=1.0\linewidth,clip=,angle=0}}
\caption{Marginal mean maps ({\it top}) and r.m.s. ({\it bottom}) maps in Galactic coordinates. From left to right, the panels show the CMB, the foreground amplitude and the foreground spectral index, respectively. The signal model and Galactic mask is the same as used by \cite{eriksen:2008b}; see text for details. The Kp2-based mask is indicated in the grey shaded region of the CMB map. The color scale in the CMB maps is linear in units of $\mu$K while the foreground amplitude is histogram equalized and in units of $\mu$K normalized to 23~GHz. The new version of the \commander~code allows the foreground parameters to be sampled inside the mask.}
\label{fig:wmap5_maps}
\end{figure*}

The FDS99 dust template amplitude is $0.949\pm 0.003$ relative to 94~GHz. This is in good agreement with the {\it WMAP} 3-yr results and indicates that the dust model of FDS99 is a remarkably accurate prediction when averaging over the high latitude sky. Note that the formal error is likely to be under-estimated, since it is based on the contribution from the noise model ($2~\mu$K white noise) and does not include model-dependent uncertainties. The CMB map and low frequency foreground amplitude and power-law maps are shown in Fig.~\ref{fig:wmap5_maps}. By eye, the maps look similar to those produced in the 3-yr analysis. The error in the CMB map is dominated by the dipole aligned with the Galactic mask (dipole X) while the foreground errors increase at lower latitudes due to the increased foreground levels.

A recent improvement in the \commander~code is the ability to fit for foreground parameters inside the Galactic mask, without affecting the CMB power spectrum. In masked regions, the large-scale CMB power and any templates (foregrounds, monopoles/dipoles) are subtracted first, based on the fits of the unmasked regions. The residual is then fitted by any remaining foreground parameters on a pixel-by-pixel basis. In this case, the low frequency power-law (amplitude and spectral index) is fitted to each pixel inside the mask. The result is a full-sky map of the foregrounds (Fig.~\ref{fig:wmap5_maps}). Except for a small discontinuity between at the mask boundary, the spectral index map is within the expected range. The dominance of free-free emission, with a theoretical spectral index $\beta_{\rm ff}\approx -2.1$, is apparent in the Galactic plane and in well-known star-forming regions such as Orion ($(l,b)\sim (210^{\circ},-10^{\circ})$), the Gum nebula ($(l,b)\sim(260^{\circ},-10^{\circ}$) and Ophiucus ($(l,b)\sim(5^{\circ},+25^{\circ})$. The average spectral index for the whole sky is $\beta=-2.90\pm0.26$. At high latitudes ($|b|>20^{\circ}$), the spectral index has a range of $-3.8<\beta<-2.1$ with an average of $\beta=-2.97\pm0.21$. The average is therefore very close to the prior mean but with a standard deviation significantly less, suggesting that the actual variation in $\beta$ is less than the standard deviation of the prior, $\Delta \beta=0.3$. In \S\ref{sec:gauss_prior_sd} we will explicitly test the sensitivity of the results on the prior.

\subsection{Monopoles and Dipoles}

An important result of the {\it WMAP} 3-yr analysis of \cite{eriksen:2008b} was the discovery of significant residual monopoles and dipoles in {\it WMAP} maps. This was later confirmed to be due to an error in the processing of the maps when estimating the offset in each map based on a model of the foregrounds at the Galactic poles \citep{hinshaw:2009}. The best-fit monopoles and dipoles in the {\it WMAP} 5-yr data are given in Table~\ref{tab:wmap5_monodipoles}. The magnitude of the monopoles and dipoles is clearly smaller than those found in {\it WMAP} 3-yr data indicating the improved estimation of these terms. However, there still may be small monopole/dipole residuals in the {\it WMAP} 5-yr data. In particular, the monopoles appear to be positively biased at $\approx 3~\mu$K in four of the five bands, while at Q-band the monopole is $5.7\pm 0.5~\mu$K. The errors are purely statistical, based on the $2\mu$K white noise added to the maps, but there will be additional sources of error from unaccounted instrumental noise and more importantly, modeling errors. It is possible for example, for power to alias from the foregrounds to the monopoles/dipoles and vice versa. We reduce this effect by applying an additional constraint on the orthogonality between these two components as described in \cite{eriksen:2008a}. However, with such a simple model of a single power-law at each pixel, which accounts for several Galactic components (synchrotron, free-free and possibly anomalous dust emission), there will be additional modeling errors. We will test the stability of these quantities by making adjustments to the model (masks, priors, templates etc.) and observing the spread in these values.
\begin{table}
\centering
 \caption{Mean and r.m.s. of monopole and dipole posteriors in the {\it WMAP} 5-yr maps. {\it Top}: For the basic model using 1 ``index'' template (see \S\ref{sec:wmap5_basic}). {\it Bottom}: For the same model but with 2 ``index'' templates, including the \cite{finkbeiner:2003} H$\alpha$ template (see \S\ref{sec:2ind}).}
\label{tab:wmap5_monodipoles}
  \begin{tabular}{lcccc}
    \hline
Band           &Monopole        &Dipole X      & Dipole Y   & Dipole Z   \\
               &($\mu$K)        &($\mu$K)      &($\mu$K)    &($\mu$K)    \\
\hline
K-band         &~$3.2\pm0.5$   &~$2.3\pm1.2$  &$\phm{-}0.4\pm0.8$    &$-1.1\pm 0.1$ \\
Ka-band        &~$3.6\pm0.5$   &~$1.0\pm1.2$  &$-2.6\pm0.8$          &$\phm{-}0.5\pm 0.1$ \\
Q-band         &~$5.7\pm0.5$   &~$1.9\pm1.2$  &$-2.1\pm0.8$          &$\phm{-}0.8\pm 0.1$ \\
V-band         &~$2.6\pm0.5$   &~$2.4\pm1.2$  &$\phm{-}0.6\pm0.8$    &$-1.3\pm 0.1$ \\
W-band         &~$2.6\pm0.5$   &~$2.1\pm1.2$  &$\phm{-}0.9\pm0.8$    &$-1.6\pm 0.1$ \\
\hline
K-band         &$\phm{-}0.5\pm0.5$   &~$2.9\pm1.2$  &$\phm{-}1.2\pm0.8$    &$-0.8\pm 0.1$ \\
Ka-band        &$\phm{-}0.3\pm0.5$   &~$2.0\pm1.2$  &$-1.3\pm0.8$          &$\phm{-}0.8\pm 0.1$ \\
Q-band         &$\phm{-}2.6\pm0.5$   &~$2.8\pm1.2$  &$-0.7\pm0.8$          &$\phm{-}1.1\pm 0.1$ \\
V-band         &$-0.0\pm0.5$         &~$3.0\pm1.2$  &$\phm{-}1.6\pm0.8$    &$-1.0\pm 0.1$ \\
W-band         &$-0.0\pm0.5$         &~$2.7\pm1.2$  &$\phm{-}1.6\pm0.8$    &$-1.4\pm 0.1$ \\
\hline      
 \end{tabular}
\end{table}

\begin{figure}[!h]
\begin{center}
\mbox{\epsfig{figure=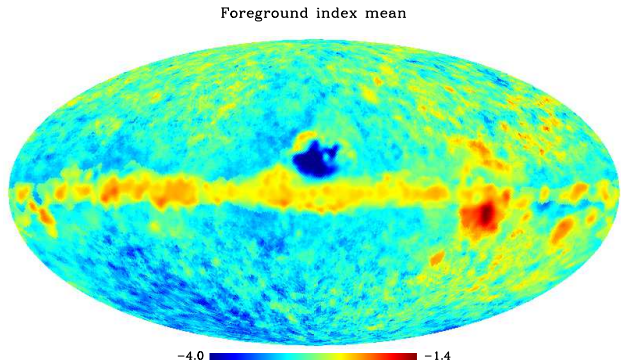,width=0.6\linewidth,clip=,angle=90}}
\caption{Low-frequency foreground spectral index mean map for {\it WMAP} 5-yr data when not fitting for monopoles and dipoles. }
\label{fig:index_nomd}
\end{center}
\end{figure}

To see how important the monopole and dipole terms are, we repeated the above analysis but not fitting for any monopoles and dipoles i.e. assume they are zero. The most obvious effect is in the foreground spectral index map, shown in Fig.~\ref{fig:index_nomd}. Besides for differences inside the masked region (particularly in the Galactic center region), there is a large-scale gradient running across the map, which is unlikely to be real, and is just a consequence of residual monopoles and dipoles in the {\it WMAP} 5-yr maps. To verify this conclusion a {\it WMAP} simulation was made, with and without the best-fitting monopoles and dipoles. The spectral index map was seen to have a similar gradient. This is also observed to some degree in the spectral index maps derived by \cite{gold:2009} using a MCMC algorithm, presumably a result of not accounting for residual monopoles/dipoles (see \S\ref{sec:index_maps} and Fig.~\ref{fig:index_maps}).

\section{{\it WMAP}5 analysis with varying models and assumptions}
\label{sec:varying_models}

\subsection{Galactic masks}
\label{sec:masks}

\begin{figure}[!h]
\begin{center}
\mbox{\epsfig{figure=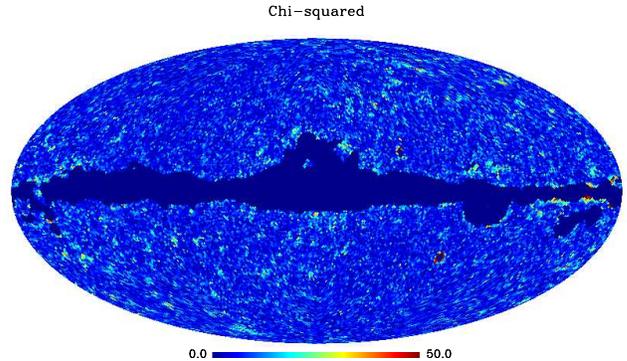,width=0.6\linewidth,clip=,angle=90}}
\caption{Map of $\chi^2$ for the basic {\it WMAP} 5-yr analysis. Most pixels are at $\chi^2<15$ and are consistent with the model, given the assumed errors. There is a slight increase in the $\chi^2$ in the ecliptic plane due to the additional unmodeled instrumental noise. A few pixels close to the Galactic plane have $\chi^2>50$, with one pixel at $\chi^2=138$.} 
\label{fig:wmap5_basic_chisq}
\end{center}
\end{figure}

The Galactic mask is extremely important for obtaining the optimal CMB estimates by masking out bright regions dominated by foreground emission (which becomes increasingly complex, particularly at low Galactic latitudes) but still allowing the maximum sky coverage (to reduce cosmic variance). From the analysis with the Kp2 mask, the $\chi^2$ map gives a handle on how good the signal model and where it is a poor fit. Fig.~\ref{fig:wmap5_basic_chisq} shows the map of $\chi^2$. Most pixels have $\chi^2<15$, which indicates a model consistent with the data and the errors; statistically, a value of 15 corresponds to a model that is rejected at the $99\%$ confidence level. There is a small increase in $\chi^2$ along the ecliptic plane, due to the additional unmodeled instrumental noise, although this is an improvement over the {\it WMAP} 3-yr results due to the additional data. A few pixels have particularly large $\chi^2$ values ($\chi^2>50$). An obvious mask choice would be to only include pixels with a $\chi^2$ below some limit. We tried a mask for $\chi^2<30$ and found that the CMB power spectrum remained unchanged.

\begin{figure}[!h]
\begin{center}
\mbox{\epsfig{figure=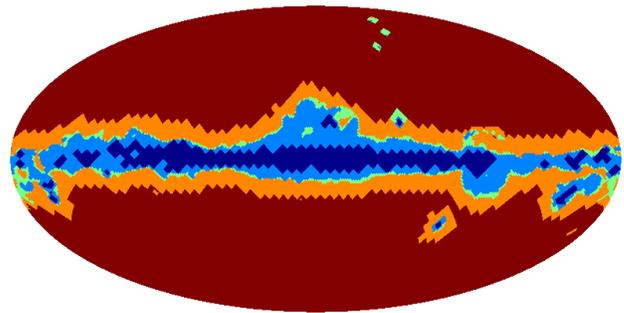,width=0.6\linewidth,clip=,angle=90}}
\caption{Map showing the coverage of Galactic masks used. The masks are Kp2 (light blue), a
$10^{\circ}$ smoothed Kp2 (orange), {\it WMAP}5
KQ85 temperature mask (green) and the {\it WMAP}5 temperature processing mask (dark blue).}
\label{fig:masks}
\end{center}
\end{figure}

We explore the sensitivity of the derived CMB and foreground
parameters by first using the exact same signal model as in
\S\ref{sec:wmap5_basic} but employing five different masks. The masks
are  i) Kp2 mask ($14.4\%$) as used in \S\ref{sec:wmap5_basic}, ii) a
$10^{\circ}$ smoothed Kp2 mask ($31.4\%$), iii) {\it WMAP}5
KQ85 temperature mask ($17.5\%$), iv) {\it WMAP}5 temperature processing mask ($6.5\%$) and v) no mask (full-sky). The coverage of each mask is shown in Fig.~\ref{fig:masks}.

\begin{figure*}[]
\begin{center}
\mbox{\epsfig{figure=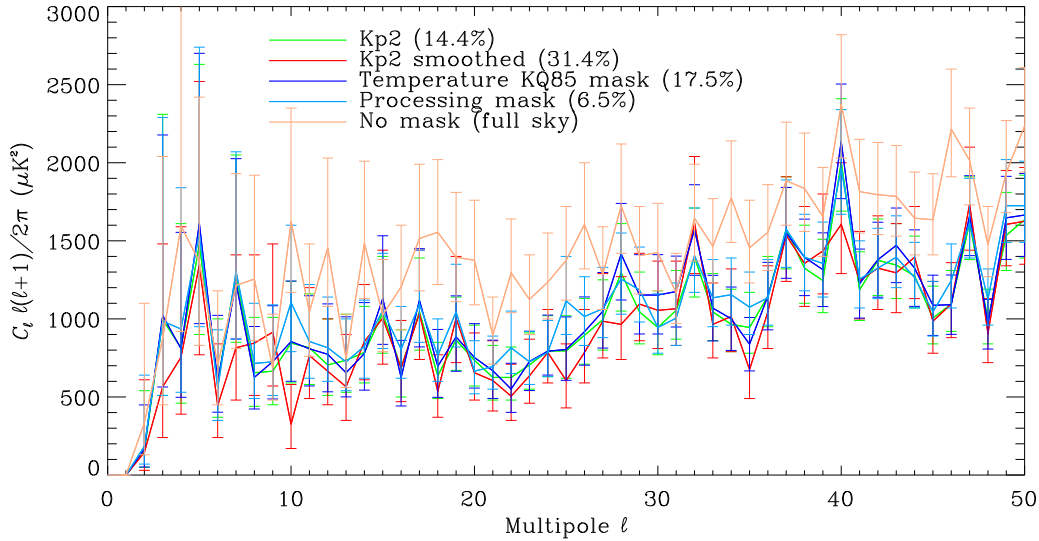,width=0.8\linewidth,clip=,angle=0}}
\caption{CMB temperature power spectra for the {\it WMAP} 5-yr data employing different Galactic masks (percentage values represent the sky area masked out).}
\label{fig:cls_masks}
\end{center}
\end{figure*}

Fig.~\ref{fig:cls_masks} shows the CMB temperature power spectra for
each mask. The signal model is the same in each case. For full-sky (no
mask) there are significant foreground contributions across the entire
$\ell$-range considered here ($\ell=2-50$) with an excess power of
$\approx 500~\mu$K$^2$. With the {\it WMAP} processing mask (6.5\%
masked out), the spectrum is quite close to the spectra employing
larger masks. However, there is some positive bias at some
$\ell$-values (e.g. $\ell= 22-27$), indicating that for this model,
this mask is not adequate to mask out the Galaxy and thus have a
negligible effect on the power spectrum. The temperature KQ85 mask and
Kp2 mask power spectra are in excellent agreement. It is interesting
to see that the largest mask, the smoothed Kp2 mask, produces a power
spectrum that is slightly lower than the others, particularly at a few specific $\ell$-values
(e.g. $\ell=10,\ell=25-28$). This is partly attributed to the approximate Gaussianized Blackwell-Rao estimator of \cite{rudjord:2008}, due to the extra correlations introduced by larger masks ($\gtrsim 30\%$ cut). Even
so, inspection of the marginal $C_\ell$ histograms does confirm that
a few specific $\ell$-values (particularly $\ell=10$) are lower, but no more than $2\sigma$, than for the larger mask, and are likely to be due to residual
foreground contamination at low Galactic latitudes. We checked that the impact on the cosmological parameters is negligible ($<<1\sigma$ for all parameters).

\subsection{H$\alpha$ templates}
\label{sec:ha_templates}

A major issue with the signal model used so far is that the majority of the foreground emission is fitted by a single power-law at each pixel. This is a broad assumption given that we know that there are at least two distinct Galactic components emitting at the lower frequencies: synchrotron radiation (which has a spatially varying spectral index) and free-free emission (which as a flatter spectral index but varies little with position). There is also substantial evidence for an anomalous dust-correlated components, possibly due to spinning dust grains, which is strong in the lower {\it WMAP} bands. The anomalous component appears to have a spectral index of $\beta \approx -3.1$ between 20 and 40~GHz \citep{davies:2006} thus one might expect a single power-law to be adequate for both components. However, the free-free emission has a spectral index $\beta_{\rm ff} \approx -2.15$ at these frequencies \citep{dickinson:2003}. If the free-free emission is significant in the higher {\it WMAP} bands, the power-law model will clearly not be a good model and the CMB estimates could be affected. Fortunately, there are full-sky free-free templates available, based on H$\alpha$ surveys \citep{dickinson:2003,finkbeiner:2003}. These are expected to be good tracers of the free-free emission for most of the sky, with the exception of low latitudes where dust absorption will affect the H$\alpha$ intensity.

\subsubsection{H$\alpha$ fixed index template}
\label{sec:2ind}

We included a free-free template based on H$\alpha$ \citep{finkbeiner:2003,dickinson:2003} as an additional template with a fixed spectral index $\beta_{\rm ff}=-2.15$. This was not included in the analysis of \cite{eriksen:2008b} because simulations showed that with this model, for the {\it WMAP} frequency range, the code did not reproduce the exact input values. In particular, the free-free amplitude was underestimated by $\approx 10\%$ (depending on the typical assumed synchrotron spectral indices) due to aliasing between the power-law component and the free-free template. This is due to a degeneracy between synchrotron and free-free
emission over the {\it WMAP} frequencies: With effectively only three available
frequencies (K,Ka,Q) and two free components (synchrotron and free-free), the resulting joint posterior
becomes strongly non-Gaussian and asymmetric. As a result, the
corresponding univariate marginal means are no longer unbiased
estimators, and not useful as point estimators. For an explicit example
of this effect, see \citet{eriksen:2007}. In the simulation, the other components (dust template amplitude, monopoles and dipoles) were reproduced correctly and more importantly, the CMB power spectra estimates were still consistent within a small fraction of the formal errors, giving us confidence that this could still improve the foreground subtraction. Additional bands, particularly at lower frequencies, can improve this situation.

When applied to {\it WMAP} 5-yr data, the foreground template amplitudes were $0.998\pm0.003$ for the FDS99 dust template at 94~GHz and $6.30\pm0.08~\mu$K~$R^{-1}$ for the Finkbeiner H$\alpha$ template, relative to 23~GHz. Similar results were obtained when using the \cite{dickinson:2003} template with a H$\alpha$ amplitude of $5.81\pm0.05~\mu$K~$R^{-1}$, relative to 23~GHz. The H$\alpha$ amplitude is below what is expected ($11.4~\mu$K~$R^{-1}$) for an electron temperature $T_{e}\approx 8000$~K, but is consistent with previous analyses \citep{davies:2006,gold:2009}. The monopoles and dipoles are given in Table~\ref{tab:wmap5_monodipoles}. It is interesting to see that the monopoles are now all consistent with 0, except for Q-band, which appears to still have a positive residual monopole of $2.6\pm0.5~\mu$K. This may be due to the improved foreground modeling while the dipoles are of a similar magnitude. This is supported by a $\chi^2=301555\pm9$ compared to $\chi^2=301667\pm9$ with just one additional degree of freedom. The residual dipoles still remain at the $\approx 2-3~\mu$K level and may not be totally negligible.

\begin{figure}[!h]
\mbox{\epsfig{figure=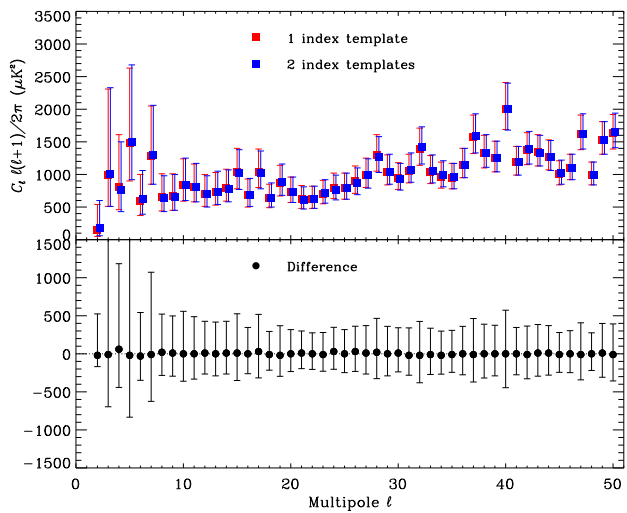,width=1.0\linewidth,clip=,angle=0}}
\caption{{\it Top}: CMB temperature power spectra for the basic {\it WMAP}5 model (1 ``index'' template) and when an additional H$\alpha$ template is included (2 ``index'' templates). {\it Bottom}: Difference power spectrum between 1 and 2 ``index'' templates.}
\label{fig:cls_1ind_2ind}
\end{figure}

\begin{figure}[!h]
\mbox{\epsfig{figure=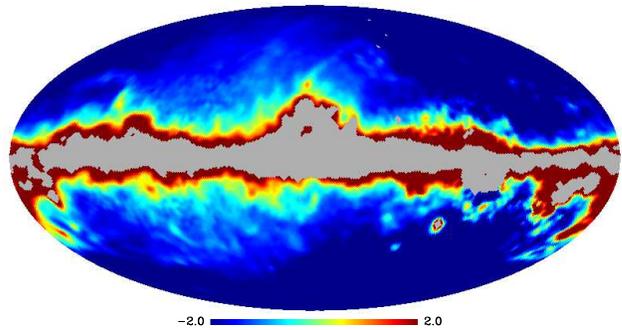,width=0.6\linewidth,clip=,angle=90}}
\caption{Difference map of the mean CMB maps from the basic {\it WMAP}5 model (1 template) and when an additional H$\alpha$ template is included (2 templates). Units are $\mu$K.}
\label{fig:cmb_map_1ind_2ind}
\end{figure}

The CMB temperature power spectra, computed with and without the additional H$\alpha$ template, are shown in Fig.~\ref{fig:cls_1ind_2ind}. The differences are negligible relative to the error bars. This tells us, assuming the H$\alpha$ map is a good template for free-free emission, that the single power-law model is an adequate model for this particular data set and the Kp2 mask. This is justified by the fact that the total level of free-free emission at high latitudes is actually quite small, relative to the CMB and other foregrounds. The H$\alpha$ intensity at high latitude is $\sim 1$~R, which corresponds to $\sim 10~\mu$K at 23~GHz and $\sim 0.5~\mu$K at 94~GHz. For comparison, at 23~GHz and at high latitudes, the foregrounds are $\sim 100~\mu$K while at 94~GHz, the signal is dominated by CMB at $\sim 30~\mu$K. The free-free component is therefore not a dominant component at {\it WMAP} frequencies, except in the Galactic plane and in regions of intense star formation. Indeed, the resulting CMB map appears (by eye) to be identical to the single template model. However, the difference map does reveal that there is a residual Galactic component, as shown in Fig.~\ref{fig:cmb_map_1ind_2ind}. This residual is $\sim 1~\mu$K for most of the sky, increasing nearer the plane to $\sim 5~\mu$K, hence why it is not visible in the CMB map itself. The structure of this residual component (Fig.~\ref{fig:cmb_map_1ind_2ind}) is similar to the H$\alpha$ map. Although it does not have an impact on the CMB power spectrum, such errors may have an impact on sensitive non-Gaussianity tests.

\begin{figure}[!h]
\mbox{\epsfig{figure=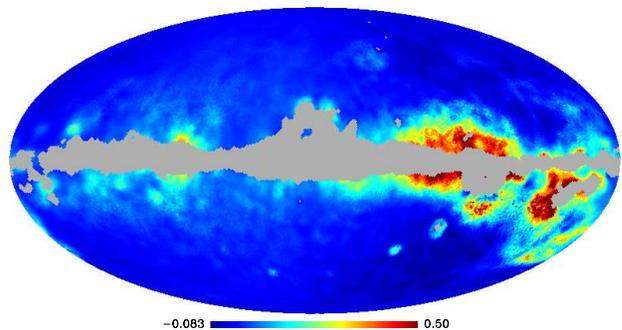,width=0.6\linewidth,clip=,angle=90}}
\caption{Difference map of the low frequency spectral indices derived from the basic {\it WMAP}5 model (1 template) and when an additional H$\alpha$ template is included (2 templates). The full range is $-0.083$ to $1.3$.}
\label{fig:index_basic-2ind}
\end{figure}

The effect on the low-frequency foreground spectral index is shown in Fig.~\ref{fig:index_basic-2ind}. For much of the sky, the differences are small ($\lesssim 0.1$), with the largest variations occurring near well-known regions of strong free-free emission such as Orion, Gum nebula and Cygnus. In these regions, spectral index decreases by $\sim 0.5$, as expected with a transition from synchrotron to free-free emission.

\subsubsection{H$\alpha$ free template}

Rather than assume the known spectral dependence for a template (the ``index'' template option), {\tt Commander} allows for an independent fit to each frequency channel (the ``free'' template option) i.e. fitting for a template amplitude coefficient for each frequency. When this option is used for the \cite{finkbeiner:2003} H$\alpha$ template, we recover the following amplitudes: 5.67, 2.72, 1.78, 0.68, 0.27 ($\pm 0.08$) $\mu$K~$R^{-1}$, at 23, 33, 41, 61 and 94~GHz, respectively. These values are slightly lower than those obtained with a fixed spectral index ($6.3 \pm 0.08~\mu$K~$R^{-1}$ at 23~GHz). The amplitudes follow a free-free spectrum very closely, with a best-fit spectral index to the amplitudes of $\beta=-2.12\pm0.05$, consistent with our assumption of $\beta_{\rm ff}=-2.15$. These coefficients are consistent with a power-law, with a slight steepening at V- and W-bands, which is actually expected for free-free emission \citep{dickinson:2003}. For this model we find no evidence for a broad bump, centered at $\sim 40-50$~GHz in the H$\alpha$ amplitudes, that has been claimed by \cite{dobler:2008b,dobler:2008c}. This may be due to this particular effect, only being visible at lower Galactic latitudes, and hence not seen with our larger mask. Indeed, we note the large variation in template coefficients when different masks are used (\S\ref{sec:templates_only}).

\subsection{Thermal dust}
\label{sec:thermal_dust}

Thermal (vibrational) dust emission comes from warm ($T_{\rm d} \sim
18$~K) dust emitting black-body radiation, peaking at $\sim 100~\mu$m
(3000~GHz). Thermal dust is a relatively weak component in most the
{\it WMAP} channels. In the lowest bands (K,Ka,Q) it is almost
negligible, while at W-band it begins to dominate over the other
components. One of the assumptions that is commonly used, in analyses of {\it WMAP} data, is that the thermal dust can be subtracted using a single FIR
template (e.g. FDS99) - either via a cross-correlation with each band,
or with a fixed spectral index. This appears to work well, and surprisingly, the amplitude is remarkably close to the that of the template i.e. $\approx 1.0$ at 94~GHz (see \S\ref{sec:wmap5_basic}). The best-fitting model of FDS99 (model
8) has an effective spectral index of $\beta_{\rm d} \approx +1.7$ at {\it
WMAP} frequencies. Dust is known to emit with a range of indices, with
a total range of $\sim 1-3$ \citep{finkbeiner:1999,dupac:2003} but
with an average of $\sim 2$ in the diffuse ISM. We now investigate the sensitivity of the results on the assumption of a spectral index for the thermal dust, and if other templates, can provide a better fit.

\subsubsection{The thermal dust spectral index}
\label{sec:dust_index}

To test the sensitivity on the assumed spectral index of the dust
template, we repeated the {\it WMAP} 5-yr analysis but varying the
assumed dust template spectral index with $\beta_{\rm
d}=+1.5,1.6,7,1.8,1.9,2.0,2.1,2.2$, covering the likely range for the
average value. To reduce the effects of free-free emission, which
could be significant at the higher frequencies, the \cite{finkbeiner:2003} H$\alpha$ template
was included.

In terms of goodness-of-fit, the total $\chi^2$ values were 302240, 301890,
301555, 301234, 300928, 300638, 300364, 300106 $(\pm 440)$, for
increasing $\beta_{\rm d}$. This is telling us that the steeper the
dust index, the better the fit. We therefore tried higher values,
$\beta_{\rm d}=2.5,3.0,4.0,5.0,6.0...10.0$. The minimum of $\chi^2$
occurs at $\beta_{\rm d}=+6.0$. This is beyond the range of plausible
physical values ($\beta_{\rm d}\approx 1-3$) and must be a consequence
of modeling errors. A possible explanation is that residuals due to
the assumption of a single power-law for the low-frequency foregrounds
(which is accounting for both synchrotron and anomalous dust emission)
are biasing the thermal dust template fit. More precisely, it is known
that the anomalous emission is tightly correlated with thermal dust
emission based on FIR data \citep{davies:2006}. If the anomalous
emission is indeed due to spinning dust, the spectrum will drop off
much more rapidly than synchrotron in the $\sim 40-100$~GHz range,
resulting in a flatter spectral index being fitted in the lower frequency channels. This in turn will
result in too much of the dust-correlated foreground being subtracted
at the higher {\it WMAP} frequencies (V- and W-bands) and thus the
dust template will prefer a steeper index. The resulting dust spectral
index ($\beta_{\rm d}=+6.0$) can, in principle, give information on
the residual and hence the spectrum of the modeled
component. However, the effective spectral index for the combined
synchrotron and spinning dust components is very sensitive on the
exact peak of the spinning dust spectrum, since it is close to the
lowest frequency channels of {\it {\it WMAP}}, in the range $\approx
20-40$~GHz \citep{draine:1998a,draine:1998b,ali-hamoud:2008}. Still, given that the
majority of spinning dust models fall off steeper than $\beta=-4$, it
is not inconceivable, that a thermal dust spectrum of $\beta_{\rm
d}=+6.0$ is compatible with such a residual. The impact on the CMB
temperature power spectrum is negligible because of the absolute level of thermal dust emission at {\it WMAP} frequencies is relatively small.

\subsubsection{Thermal dust templates}

So far, we have assumed that the FDS99 template, at 94~GHz, is the best spatial template of thermal dust at {\it WMAP} frequencies. But given the simplicity of the fitted model used by \cite{finkbeiner:1999} (2 component dust model), one can imagine that there may be some errors in the extrapolation. To see how sensitive the results were on this template, we replaced the FDS99 template with other templates. First, we tried the $E(B-V)$ reddening maps, which are temperature corrected, and are essentially proportional to the dust column density. The results were almost identical to the FDS99, with a small increase in the $\chi^2$ value; 295842 for the $E(B-V)$ case compared to 295355 in the FDS99 case. This indicates that, at least spatially, the $E(B-V)$ map and the FDS99 maps are very similar.

We also tried the raw SFD98 $100~\mu$m intensity map. For this case, $\chi^2=350863$, which is significantly higher than in previous cases. Visual inspection of the maps revealed that a few pixels, in the Orion region (near Barnard's loop), were anomalous. The CMB map was strongly negative ($-494~\mu$K) in this region, while the low frequency spectral index was at $-4.0$ i.e. pegged at the prior. Also, the power spectrum showed additional power at $\ell \lesssim 10$, indicative of foreground residuals. With a larger mask, this disappeared. As expected, the FDS99 template is more representative of the morphology of the real sky, than the raw $100~\mu$m intensity map.

\subsubsection{Free dust template}

Up to now, the FDS99 dust template has been used to account for the thermal dust contribution at higher frequencies; we fitted for a template with a fixed spectral index e.g. $\beta=+1.7$. We now relax this constraint by fitting for the same FDS99 template ``freely'' at each frequency. The most important aspect of this model, is that the FDS99 template should trace the anomalous emission at lower frequencies, as seen in \S\ref{sec:templates_only}. Several combinations were tried, such as including a H$\alpha$ template with fixed spectral index or not. The results of these were similar with no significant variations in the CMB spectrum. For a fixed spectral index H$\alpha$ template, and the Kp2 mask, the H$\alpha$ amplitude was $8.02\pm0.04~\mu$K~$R^{-1}$ at 23~GHz, similar to what was found in \S\ref{sec:templates_only} (Table~\ref{tab:template_amplitudes}). The FDS99 amplitude coefficients were 0.67, 0.08, 0.136, 0.36, 1.12 ($\pm 0.09$) $~\mu$K $~\mu$K$^{-1}$ at 23, 33, 41, 61 and 94~GHz, respectively. The thermal dust contribution appears to be robust, with an amplitude at high frequencies remains at close to the predicted level from FDS99; this is in contrast to the template only fits (\S\ref{sec:templates_only}/Table~\ref{tab:template_amplitudes}) where the values varied considerably with the exact model and mask.  The dust contribution then decreases substantially at frequencies immediately below W-band with a minimum at 33~GHz. At 23~GHz, the dust-correlated contribution then rises quickly, which is indicative of spinning dust. However, the value at 33~GHz appears to be much lower than expected for typical spinning dust models. This is likely to be due to aliasing of power between the template and the foreground captured by the power-law term in each pixel. Moreover, the FDS99 amplitude at 23~GHz is $\approx 1/10$ of the value found with template only analyses in \S\ref{sec:templates_only} and previous analyses \citep{davies:2006}. Indeed, simulations showed that the dust template coefficients were unreliable, and were very sensitive on the details of the spinning dust model. We can speculate that only a fraction of the anomalous dust signal is truly correlated with the FDS99 template and therefore the rest is accounted for by the low frequency power-law component. This would also explain why a single low frequency power-law appears to be an adequate model, at least for {\it WMAP} data, for the combination of synchrotron and anomalous dust foregrounds.



\subsubsection{Thermal dust as a power-law at each pixel}
\label{sec:thermaldust_powerlaw}

Rather than using a fixed spatial template, we relaxed this assumption by fitting for a dust amplitude in each pixel. This means fitting for two power-laws in each pixel, one with a free spectral index (for the low frequency foregrounds) and one with a fixed index (for the thermal dust). {\it WMAP} data do not allow fitting for a spectral index as well (both in terms of degrees of freedom and frequency coverage), so this has to be fixed for the dust power-law e.g. $\beta_{\rm d}=+1.7$. The first attempt looked quite reasonable, although the fitted monopoles were biased high ($4-7~\mu$K). Since the results so far indicated that residual monopoles/dipoles were small in the {\it WMAP} 5-yr data, these were fixed at zero. The CMB power spectrum was consistent with the previous results. 

\begin{figure}[!h]
\mbox{\epsfig{figure=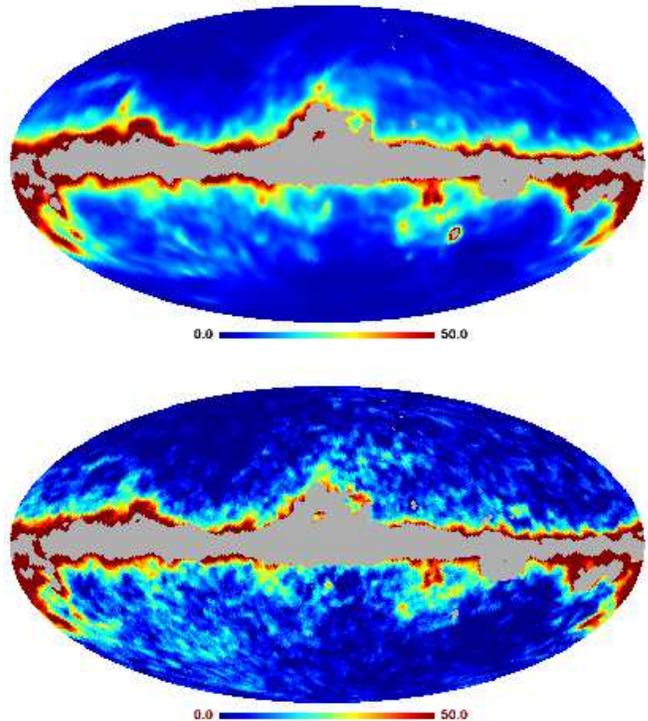,width=1.0\linewidth,clip=,angle=0}}
\caption{{\it Top}: The FDS99 thermal dust prediction template at 94~GHz. {\it Bottom}: Marginal mean map for the thermal dust amplitude (normalized at 94~GHz), when modeled as a power-law with fixed spectral index, $\beta_{\rm d}=+1.7$. }
\label{fig:dust_amp_maps}
\end{figure}

The marginal mean map for the dust amplitude is compared with the FDS99 template in Fig.~\ref{fig:dust_amp_maps}. You can see that, due to the limited frequency range of {\it WMAP}, the dust amplitude is relatively noisy at high latitudes. However, there is a good overall correspondence between the two maps; most of the brighter features are in both maps, with similar amplitudes, particularly at lower latitudes where the emission is stronger. The FDS99 does a remarkable job of predicting the thermal dust amplitude at 94~GHz. We did not see a significant improvement in the $\chi^2$ map, except inside the mask (in the Galactic plane), where the FDS99 prediction is unlikely to be accurate due to the complexities of a wider variation in dust properties in the Galaxy. Additional high frequency channels, as will be provided by the upcoming {\it Planck} satellite, will provide a much better view of the thermal dust component at frequencies $\gtrsim 100$~GHz.

\subsection{The spectral index prior}
\label{sec:gauss_prior}

An important parameter is the Bayesian prior that is imposed on the parameters, and in particular, the spectral index of the power-law component. \commander~implements two types of priors on the spectral index. The first is a uniform prior that defines the range of valid spectral indices. For the low frequency foregrounds, we typically use a range that encompasses any plausible value e.g. $-4.0<\beta<-1$. The second is a more restrictive Gaussian prior, defined by a mean spectral index and standard deviation. The advantage of such a prior is that in the high signal-to-noise regime, where the data may prefer a value several standard deviations away from the mean, the solution will move towards the preferred value since the Gaussian prior never goes completely to zero (until it hits the uniform prior). An example of this can be seen in Fig.~\ref{fig:wmap5_maps} where near the plane, $\beta \approx -2.1$, which is $3\sigma$ away from the prior mean. On the other hand, in the low signal-to-noise regime, the prior becomes more important, defining the range of values that the samples can take. It is therefore important to choose a prior that represents the best available knowledge without being too weak or too strong.

For the analyses so far, we've used a Gaussian prior\footnote{In practice, the priors need to be multiplied by the appropriate Jeffreys ignorance prior to account for the different volumes of parameter space for non-linear parameters; see \cite{eriksen:2008a} for details.} with mean $\beta=-3.0$ and standard deviation $\Delta \beta=0.3$. We now investigate what impact this assumption has on the results.

\subsubsection{Mean}
\label{sec:gauss_prior_mean}

We ran the basic model (as in \S\ref{sec:wmap5_basic}) but varying the Gaussian prior mean, $\beta$. We tried $-3.3 < \beta < -2.7$ in steps of $0.1$. The total $\chi^2$ values were 284519, 284030,283901,284157,284839, 286012, 287771 $(\pm 434)$, indicating that $\beta=-2.9$ gave a marginally better fit. The CMB temperature power spectrum remains unchanged over this range, showing the robustness of the model.

The FDS99 dust template amplitude varied from 0.978 (for $\beta=-2.7$) to 0.892 (for $\beta=-3.3$), showing that the choice of prior is relatively important for interpreting the individual foreground components. Similarly, the monopoles varied systematically as a function of $\beta$, by up to $\approx 10~\mu$K over the range of priors used here. The correlation between $\beta$ and the monopole amplitude has negligible effect on the CMB power spectrum since it only affects the overall offset in the map.

\begin{figure}[!h]
\mbox{\epsfig{figure=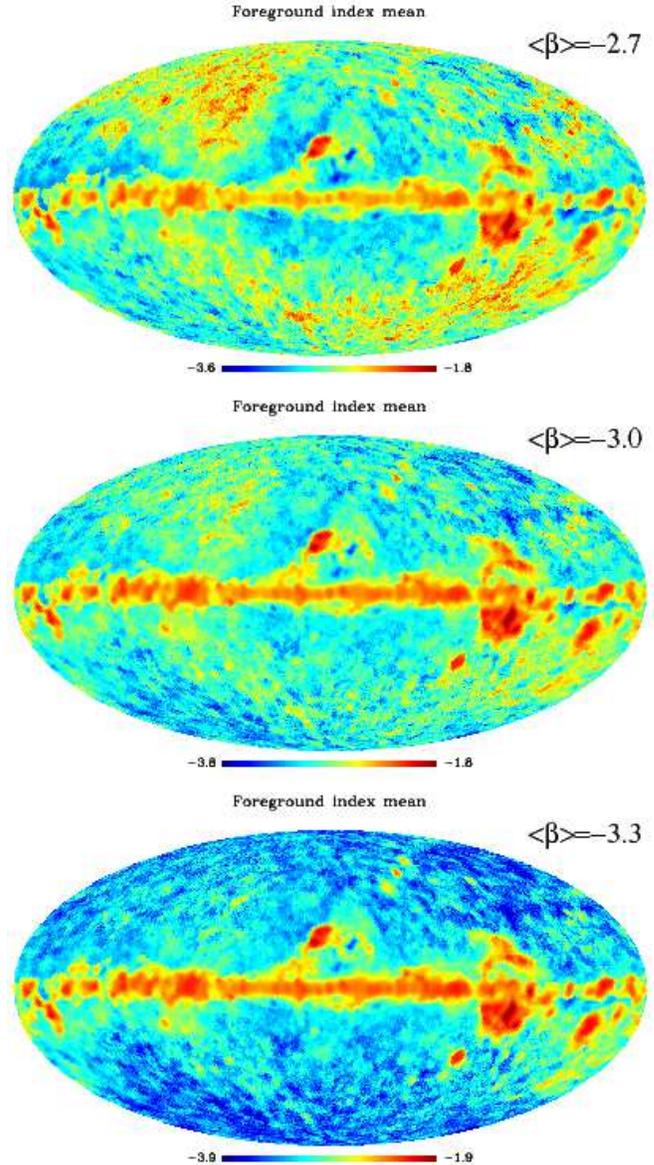,width=1.0\linewidth,clip=,angle=0}}
\caption{Low frequency foreground spectral index maps for different Gaussian prior means. From top to bottom, $\beta=-2.7,-3.0,-3.3$ with a standard deviation, $\Delta\beta=0.3$.}
\label{fig:gaussmean_index_maps}
\end{figure}

Fig.~\ref{fig:gaussmean_index_maps} shows the foreground spectral index map derived for $\beta=-2.7,-3.0,-3.3$. The bright regions of emission do not change by much ($\Delta \beta \lesssim 0.1$), showing the robustness of the result irrespective of the prior. However, at high latitudes, there is a significant difference, indicating that the prior is important in regions of weak emission. The most striking difference is in the broad $\sim 180^{\circ}$ diameter ``ring'' that is very clear for the $\beta=-2.7$ prior, evident for the $\beta=-3.0$ prior, but is not visible for the $\beta=-3.3$ prior (Fig.~\ref{fig:gaussmean_index_maps}). The ring feature is unlikely to be emanating from a particular structure in the Galaxy. It is probably a consequence of low signal-to-noise regions lying outside the ring, which will tend towards the prior mean (e.g. $\beta=-2.7$ or $\beta=-3.0$) and the steeper spectrum regions lying inside, coming from the radio spurs (the North Polar Spur) and flatter spectrum emission (e.g. free-free) near the Galactic center. When the prior mean is at $\beta=-3.3$, this matches closely the spectrum of the nearby emission making the feature largely disappear (Fig.~\ref{fig:gaussmean_index_maps}). We also note that around the Galactic center region, outside the Kp2 mask, the low frequency spectral index is {\it steeper} than average ($\beta \approx -3.3$), which is in disagreement to the hard synchrotron hypothesis to account for the ``haze'' \citep{dobler:2008a}. A similar analysis was repeated for the case with 2 templates, and similar results were found.

\subsubsection{Standard deviation}
\label{sec:gauss_prior_sd}

\begin{figure}[!h]
\mbox{\epsfig{figure=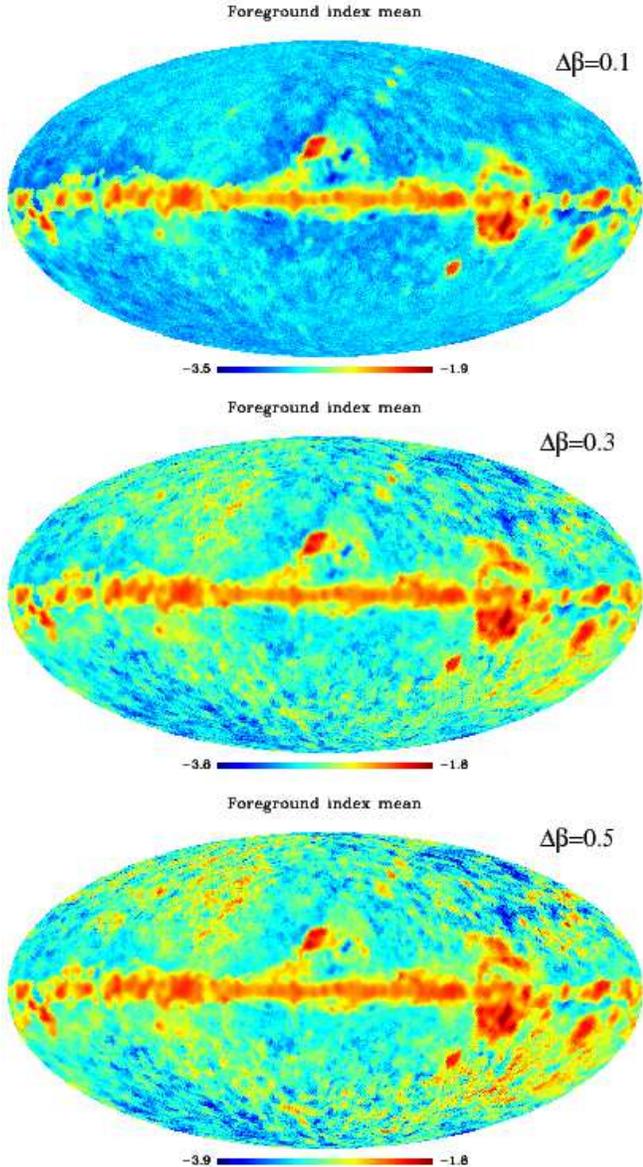,width=1.0\linewidth,clip=,angle=0}}
\caption{Low frequency foreground spectral index maps for different Gaussian prior standard deviations. From top to bottom, $\Delta \beta=0.1,0.3,0.5$ with a mean, $\beta=-3.0$.}
\label{fig:ind_prior_maps}
\end{figure}

We now vary the standard deviation, keeping the mean fixed at $\beta=-3.0$. We tried $\Delta \beta=0.1,0.2...0.5$ in steps of 0.1. The $\chi^2$ values were 313191, 289437, 284157, 282694, 282280 $(\pm 434)$. As expected, the goodness-of-fit increases steadily with increasing $\Delta \beta$, but with a significant jump for $\Delta \beta \lesssim 0.2$. This suggests that the prior is dominating for $\Delta \beta \lesssim 0.2$ and hence there are real fluctuations of $\Delta \beta \gtrsim 0.2$ on the sky. The maps of the spectral index mean for $\Delta \beta=0.1,0.3,0.5$ are shown in Fig.~\ref{fig:ind_prior_maps}. It is clear that $\Delta \beta=0.1$ strongly dampens the variation of spectral index  at high latitudes. Our original choice of $\Delta \beta=0.3$ is justified since there is only a very small change from $\Delta \beta=0.3$ to $\Delta \beta=0.5$. 
 
\subsubsection{Running of the spectral index}
\label{sec:curvature}

The assumption of a power-law, for the low frequency foregrounds, is unlikely to be accurate over a wide range of frequencies. Synchrotron emission is known to steepen with frequency due to spectral aging of the cosmic ray electrons, although multiple synchrotron components (with different spectral indices) may result in a flattening of the effective index. Anomalous emission could also modify the effective spectral index, probably resulting in a steepening at higher frequencies.
A first order correction to the power-law is to fit for an additional parameter in the exponent, usually referred to as the ``running of the index'', or curvature parameter, $C$ (see Table~\ref{tab:models}). Note that positive values of $C$ represent a flattening of the spectral index with increasing frequency, while negative values represent a steepening with frequency. 

The {\it WMAP} data alone do not allow for a reliable estimate of $C$ for each pixel. Instead, we fit a model with a fixed value of $C$ for the whole sky, relative to 23~GHz. We found that there was a slight preference for a flattening of the spectral index; the $\chi^2$ values where $295362, 293725, 292757, 292386, 292776, 294427$ ($\pm 450$), for $C=0.0, 0.1, 0.2, 0.3, 0.4, 0.5$, respectively. The best-fitting value therefore appears to be for $C=0.3$ i.e. a flattening of the spectral index at higher frequencies. This is also the preferred value found by \cite{kogut:2007} for the polarized synchrotron, although they found $C\approx-0.4$ for the unpolarized emission. The discrepancy is likely to be due to modeling errors and the additional components in temperature, that do not emit strongly in polarization (e.g. free-free, spinning dust). The flattening of the index is not what one would initially expect in temperature, since a strong spinning dust component would likely result in a negative curvature. There are several explanations that could account for this: i) multiple synchrotron components with a range of spectral indices would result in a flattening of the index at higher frequencies, ii) free-free emission unaccounted for by the H$\alpha$ template, iii) thermal dust emission not accounted for by the FDS99 template. It is interesting to note that \cite{deOliveira-Costa:2008} observed an overall flattening ($C>0$) when fitting data from 10~MHz to 100~GHz, adopting a reference frequency of 5~GHz.

\subsection{Template only fits}
\label{sec:templates_only}

The fitting of spatial templates to CMB data has been one of the most effective ways of reducing and understanding foreground emissions \citep{hinshaw:2007}. The template amplitude coefficients can also give additional insight into the scaling of the foregrounds. We investigate the performance and results of fitting templates only to the {\it WMAP} 5-yr data.

\subsubsection{Haslam 408~MHz, H$\alpha$ and FDS99}

We begin by using ``standard'' templates for 3 components: \cite{haslam:1982} 408~MHz as a tracer of synchrotron, H$\alpha$ for free-free emission and FDS99 for thermal dust (the FDS99 template also accounts for the low frequency anomalous dust-correlated emission). Each template is fitted to each frequency channel independently, using pixels outside the Galactic mask. Monopoles and dipoles were not fitted for.

The template amplitude coefficients are given in Table~\ref{tab:template_amplitudes} for different Galactic masks. It can be seen that there are significant variations in the coefficients, particularly at the higher frequencies. It is clear that strong emission just outside the mask can have a strong effect on these single-fit numbers. Overall, the numbers are similar to those found by previous authors \citep{banday:2003,bennett:2003,davies:2006}. The FDS99-correlated foreground is a minimum at V-band but increases quickly at lower frequencies. This result is one of the key arguments for an anomalous component while the spinning dust origin naturally explains the strong correlation with traditional dust emission, as traced by the FDS99 template. The spectral index of this foreground, in the {\it WMAP} frequency range, is $\beta \approx -4$ which is consistent with a spinning dust type of spectrum. The individual coefficients correspond to a synchrotron spectral index of $\beta_{\rm s} \approx -3.0$, consistent with a slight steepening relative to GHz frequencies. The 408~MHz coefficients on the other hand fall off somewhat slower than what would be expected from synchrotron emission, with $\beta_{\rm s} \approx -2.5$, probably due to aliasing of power between the templates and the low signal of synchrotron emission at the higher {\it WMAP} frequencies. The H$\alpha$ coefficients also fall off slower than expected, with $\beta \sim -1.4$, rather than the expected value for free-free emission, $\beta_{\rm ff}=-2.15$. Again, this is likely to be due correlations between the templates. Still, the H$\alpha$ coefficients are significantly lower than what is expected from theory, by almost a factor of 2. For example, at 22.8~GHz we obtain coefficients in the range $6.67-8.52~\mu$K$~R^{-1}$ at 22.8~GHz, while for a typical electron temperature of $T_e=8000$~K, the theoretical value is $11.4~\mu$K~$R^{-1}$ \citep{dickinson:2003}. The origin of this is still not understood, but is likely to be related to the correlations between the various components.

\begin{table*}
\centering
 \caption{Template amplitude coefficients for a 3 template fit of Haslam et al. 408~MHz, Finkbeiner H$\alpha$ and FDS99 94~GHz. Units are K~K$^{-1}$, $\mu$K~$R^{-1}$ and $\mu$K~($\mu$K)$^{-1}$, respectively.}
\label{tab:template_amplitudes}
  \begin{tabular}{ll|cccc}
    \hline
Template  &Band    &\multicolumn{4}{c}{Mask}              \\
          &        &KQ85             &KQ75             &Kp0                &Kp2                 \\
\hline 
408~MHz &K         &$4.34\pm0.08$    &$4.24\pm0.11$    &$4.43\pm0.13$      &$4.57\pm0.08$   \\
        &Ka        &$1.67\pm0.08$    &$1.80\pm0.11$    &$1.88\pm0.13$      &$1.81\pm0.08$   \\
        &Q         &$1.02\pm0.08$    &$1.20\pm0.11$    &$1.26\pm0.13$      &$1.15\pm0.08$   \\
        &V         &$0.42\pm0.08$    &$0.64\pm0.11$    &$0.69\pm0.13$      &$0.55\pm0.08$   \\
        &W         &$0.25\pm0.08$    &$0.48\pm0.11$    &$0.52\pm0.13$      &$0.37\pm0.08$   \\
\hline
H$\alpha$ &K       &$6.67\pm0.09$    &$6.76\pm0.16$    &$7.38\pm0.14$      &$8.52\pm0.09$   \\
          &Ka      &$3.87\pm0.09$    &$4.62\pm0.16$    &$4.49\pm0.14$      &$5.20\pm0.09$   \\
          &Q       &$2.82\pm0.09$    &$3.67\pm0.16$    &$3.37\pm0.14$      &$4.20\pm0.09$   \\
          &V       &$1.62\pm0.09$    &$2.50\pm0.16$    &$2.11\pm0.14$      &$2.74\pm0.09$   \\
          &W       &$1.19\pm0.09$    &$2.06\pm0.16$    &$1.64\pm0.14$      &$2.26\pm0.09$   \\
\hline
FDS99~94GHz  &K      &$6.03\pm0.11$    &$5.64\pm0.11$    &$5.58\pm0.16$      &$5.62\pm0.09$   \\
           &Ka     &$1.79\pm0.11$    &$1.15\pm0.11$    &$1.23\pm0.16$      &$1.33\pm0.09$   \\
           &Q      &$0.81\pm0.11$    &$0.11\pm0.11$    &$0.22\pm0.16$      &$0.30\pm0.09$   \\
           &V      &$0.36\pm0.11$    &$-0.36\pm0.11$   &$-0.25\pm0.16$     &$-0.18\pm0.09$  \\
           &W      &$0.89\pm0.11$    &$0.13\pm0.11$    &$0.26\pm0.16$      &$0.34\pm0.09$   \\
\hline      
\end{tabular}
\end{table*}

\begin{figure}[!h]
\mbox{\epsfig{figure=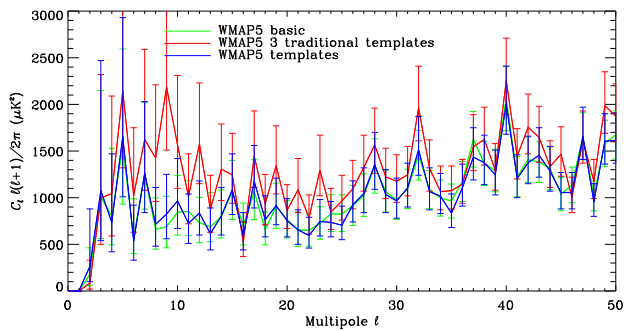,width=1.0\linewidth,clip=,angle=0}}
\caption{CMB temperature power spectra from \commander~ including the basic model, 3 traditional templates (408~MHz, H$\alpha$, FDS99~94~GHz) and 3 {\it WMAP} templates (K-Ka, H$\alpha$, FDS99~94~GHz). All analyses are using the KQ85 temperature mask.}
\label{fig:cls_3freetemplates}
\end{figure}

The CMB power spectrum from this template analysis and for the Kp2 mask is shown in Fig.~\ref{fig:cls_3freetemplates}. The power spectrum is clearly strong contaminated by foreground emission with a significant positive bias, particularly at $\ell <20$. This is largely because the synchrotron emission at 23~GHz is not perfectly traced by the 408~MHz map. Spatial variations in the synchrotron spectral index will effect the morphology of the map resulting in an imperfect template. Moreover, the 408~MHz map is known to have baseline offsets resulting in stripes in the final maps. Finally, there are residual extragalactic sources that have not been accounted for, although at the working resolution of $3^{\circ}$ the impact of these residuals is expected to be small. 

\subsubsection{K-Ka template}

As shown above, using the ``standard'' templates does not fully remove all the foreground emissions on large angular scales. A better synchrotron template, proposed by the {\it WMAP} team, is the K$-$Ka difference map \citep{hinshaw:2007}. This has the advantage of perfectly canceling out the CMB and thus containing only foreground emission at an effective frequency within the {\it WMAP} frequency range. Furthermore, the absolute calibration will be perfectly matched to the other {\it WMAP} frequencies. The disadvantages are i) the K-band and Ka-band maps cannot be used for cosmological analyses and ii) the template amplitude coefficients are more difficult to quantify, since they are effectively a contribution of at least two (or more) Galactic components.

We analyzed the {\it WMAP}5 data using the K$-$Ka difference template, fixing the dust spectrum to $\beta_{\rm d}=+1.7$ and the free-free spectrum to $\beta_{\rm ff}=-2.15$. The corresponding CMB power spectrum, for the KQ85 mask, is compared with the traditional templates in Fig.~\ref{fig:cls_3freetemplates}. The power spectrum is very close to the basic {\it WMAP}5 analysis (\S\ref{sec:wmap5_basic}) indicating the improvement in using the K$-$Ka map over the Haslam et al. (1982) 408~MHz. The improvement is partly due to the more accurate low frequency template, but also the fact that the K- and Ka-bands can no longer be used because these data are the bases of the template. The template amplitude coefficients are given in Table~\ref{tab:wmap_template_amplitudes}. The numbers are close to the {\it WMAP} values from \cite{hinshaw:2007} but not exactly the same. The differences are likely to be due to the lower resolution of our analysis ($3^{\circ}$ compared to $1^{\circ}$) and the modification of the temperature mask due to the additional smoothing used here. We also tried fitting of monopoles and dipoles with similar results.

\begin{table}
\centering
 \caption{Template amplitude coefficients for an analysis with Q-, V- and W-bands adopting the K-Ka difference map as a low frequency template. Units are  K~K$^{-1}$, $\mu$K~$R^{-1}$ and $\mu$K~($\mu$K)$^{-1}$, for the K-Ka, H$\alpha$ and FDS99~94~GHz templates, respectively. The KQ85 mask was adopted.}
\label{tab:wmap_template_amplitudes}
  \begin{tabular}{lccc}
    \hline
Band      &K-Ka               &H$\alpha$        &FDS99 94~GHz        \\
          &                   &                 &                  \\
\hline 
Q         &$327\pm14$    &$1.21\pm0.11$    &$1.21\pm0.11$    \\
V         &$115\pm14$    &$0.60\pm0.11$    &$0.60\pm0.11$     \\
W         &$50\pm14$  &$0.35\pm0.11$    &$0.35\pm0.11$     \\
\hline      
\end{tabular}
\end{table}


\subsection{CMB pixel fitting}
\label{sec:cmb_pixel_fitting}

The results from the previous sections included full-sky foreground
amplitude and spectral index maps where the low frequency foreground
is fitted inside the mask. One issue when interpreting the foreground
results is that only the large scale CMB fluctuations are constrained
within the mask, resulting in CMB residuals inside the masked area
that can effect the foreground reconstruction. This is the reason for
the discontinuity between the masked and unmasked region, as seen in
Fig.~\ref{fig:wmap5_maps}. To obtain a more accurate foreground map,
\commander~can also fit for the CMB in the pixel domain. In this model
the CMB is simply fitted by an amplitude in each pixel, since the CMB
spectrum is known (Table~\ref{tab:models}); this is similar to the MCMC algorithm described by
\cite{eriksen:2006}. Each pixel is treated as independent, therefore
spatial correlations are not taken into account, which can result in a
loss of information and hence noisy solution at small angular
scales. 

\begin{figure*}[]
\begin{center}
\mbox{\epsfig{figure=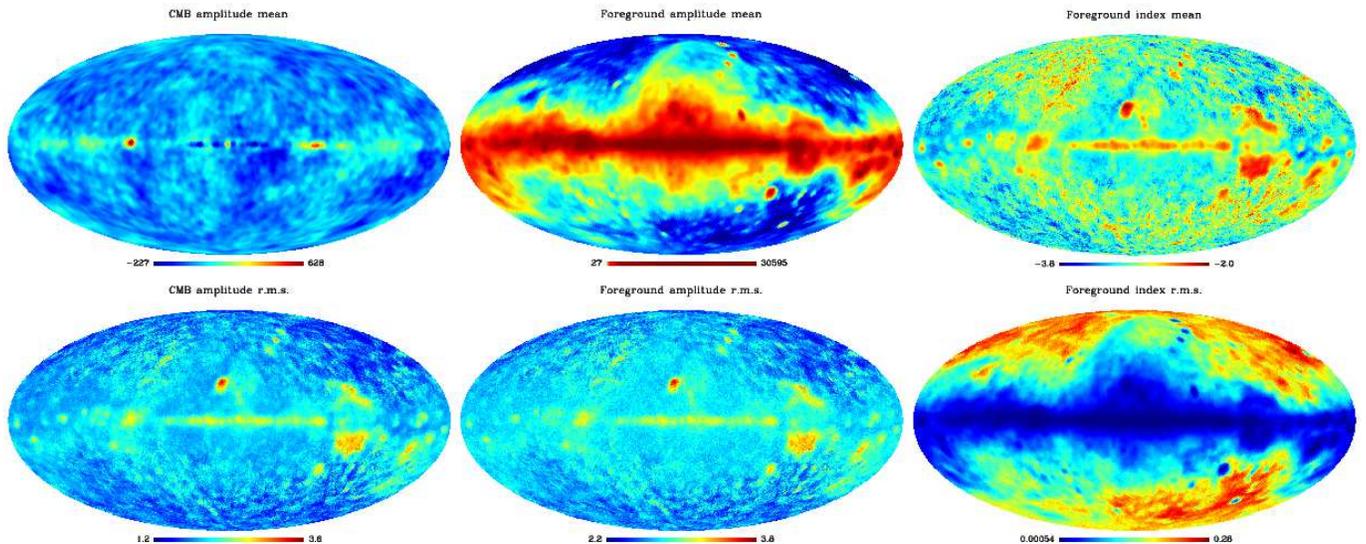,width=1.0\linewidth,clip=,angle=0}}
\caption{Reconstructed mean and r.m.s. maps for the CMB, low frequency foreground amplitude and spectral index. Here, no mask was used and the CMB is fitted pixel-by-pixel with an amplitude. }
\label{fig:cmb_pixel_maps}
\end{center}
\end{figure*}

Fig.~\ref{fig:cmb_pixel_maps} shows the margin mean maps for
the CMB, foreground amplitude and index, and their associated
r.m.s. uncertainties. For most of the sky, the results are very
similar to the results already presented. However, there are
significant differences inside the mask due to the different treatment
of the CMB in these pixels. The CMB map reveals where the model/frequency range is not adequate. First, the strong free-free regions are seen as bright
pixels in the CMB map (e.g. Cygnus at $l\approx 90^{\circ}$) due to
the flatter free-free spectral index. In the Galactic center region,
there is a slight lack of power. Examination of the spectrum at these
pixels reveals nothing special, except for a flattening at W-band due
to thermal dust. The likely reason for this is the mis-scaling of the
FDS99 94~GHz dust template at low latitudes. In the Galactic plane, there is a
complicated mix of dust properties (composition, temperatures,
emissivities etc.) which is unlikely to be well-represented by the
simple 2-component fit made by \cite{finkbeiner:1999}.

Rather than using a fixed spatial template, it is also possible to fit
for the thermal dust component as an additional foreground to be
fitted in each pixel. Clearly there are not enough channels to fit for
a more complicated model. What can be done, is to fit for the thermal dust
amplitude with a fixed spectral index (as in \S\ref{sec:thermaldust_powerlaw}). This reduced the negative values observed near the Galactic Center. We found
that $\beta_{\rm d}=2.0$ gave a better fit with $\chi^2=$1804915
compared to $\chi^2=1885888$ for $\beta_{\rm d}=+1.7$. As we saw in \S\ref{sec:dust_index}, but in a different context, the preferred value is actually
steeper than this and reflects the unmodeled flatter spectral index
(e.g. free-free) emission that is biasing the low frequency spectral
index. The monopoles and dipoles for this model are in good agreement
with the previous models, although they are statistically much more
accurate (ignoring modeling errors). From low to high frequency they
are: $0.177\pm0.003$, $-1.844\pm0.011$, $3.188\pm0.007$,
$-1.930\pm0.011$ and $0.408\pm0.003~\mu$K.

\section{Comparison of CMB and foregrounds with other works}
\label{sec:comparison}

\subsection{CMB maps and low-l power spectra}

In this paper, we have produced several versions of the CMB map with different assumptions on the foreground signal model. We now make a more detailed comparison between these maps and those available in the literature. We take our CMB map from the 2 template plus power-law model (\S\ref{sec:2ind}) and compare it with i) the {\it WMAP} 5-yr ILC map \cite{hinshaw:2009}, ii) the Harmonic ILC map of \cite{kim:2008} and iii) the {\it WMAP} 5-yr MCMC-derived best-fit CMB map of \cite{gold:2009}. The maps were smoothed to a common $3^{\circ}$ resolution and downgraded to $N_{\rm side}=64$. The MCMC map is already at $N_{\rm side}=64$ thus we simply smooth to $3^{\circ}$ resolution.

\begin{figure}[!h]
\mbox{\epsfig{figure=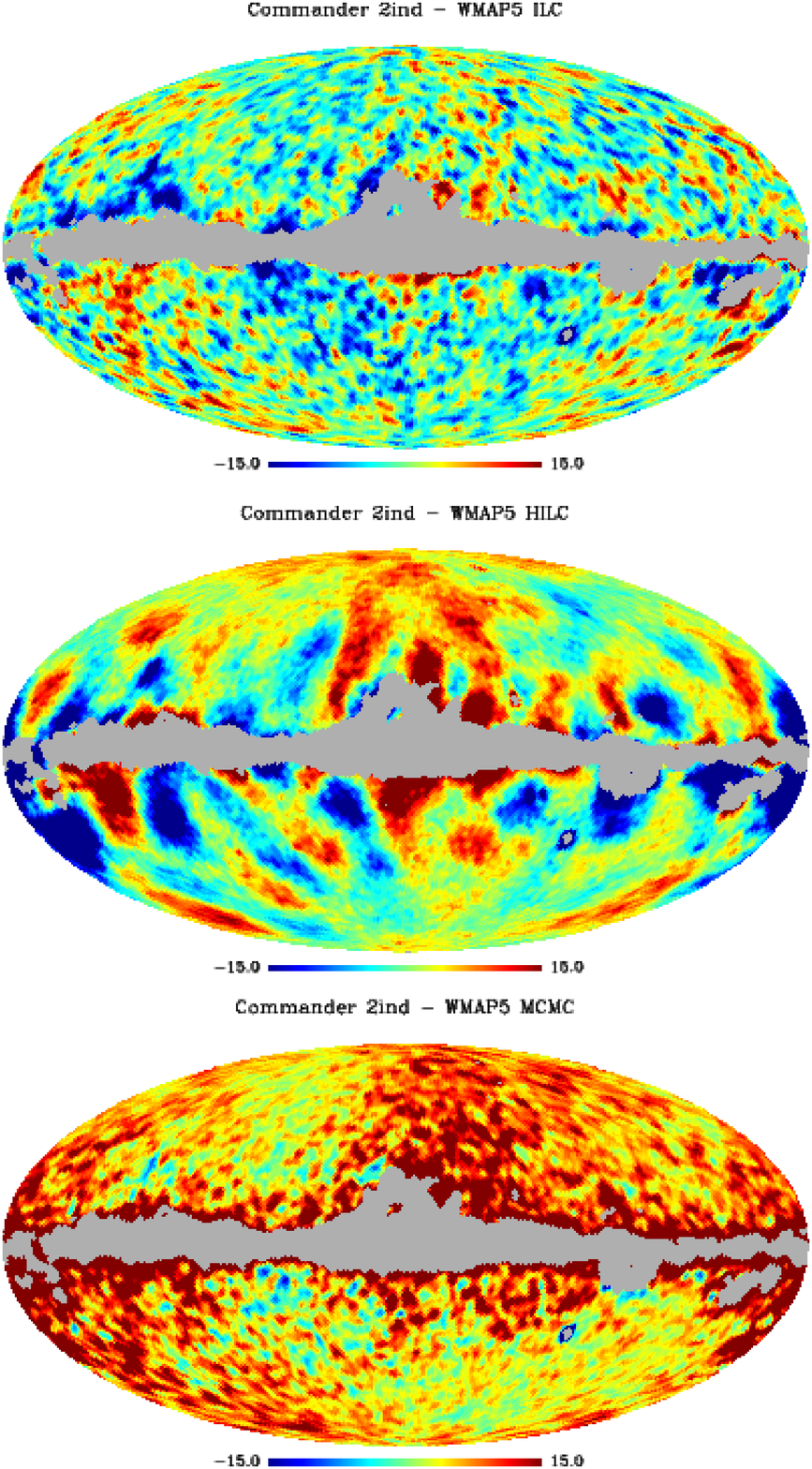,width=1.0\linewidth,clip=,angle=0}}
\caption{CMB temperature difference maps, relative to the \commander~2 index template plus power-law model. All maps have been smoothed to $3^{\circ}$ resolution and are displayed with a linear scale ranging from $-15~\mu$K to $+15~\mu$K. {\it Top}: {\it WMAP} 5-yr ILC map; {\it Middle}: Harmonic ILC map; {\it Bottom}: {\it WMAP} 5-yr MCMC map. }
\label{fig:cmb_diff_maps}
\end{figure}

A by eye visual inspection of the CMB temperature maps reveals no obvious differences, except in the Galactic plane, which is clearly strongly contaminated by foregrounds. To make a more accurate comparison, we mask the Galactic plane and make difference maps as shown in Fig.~\ref{fig:cmb_diff_maps}.

There is good overall agreement with the {\it WMAP} ILC (WILC) map with very little large scale power. There are a few regions close to the Galactic plane mask and LMC that are clearly different, with more power in the ILC map (blue regions on Fig.~\ref{fig:cmb_diff_maps}). This to be expected due to the extra degrees of freedom afforded by the power-law at each pixel, compared to the weights of the ILC method which are computed over large regions. The most apparent difference is the significant amount of small-scale power, which is a consequence of the bias correction term employed in the WILC map, resulting in significant ``noise'' at scales of $\sim 1^{\circ}$.

However, there appears to be significant differences, even on large angular scales, between the \commander~map and the Harmonic ILC map. Indeed, \cite{kim:2008} noted that most of the low-$\ell$ powers of the HILC map are higher than those of the WILC map. This suggests that the HILC map contains some residual foreground power, as seen in Fig.~\ref{fig:cmb_diff_maps}.

The {\it WMAP} MCMC-derived map appears to show a significant positive bias. Outside the mask, the MCMC map has a mean of $-8.6~\mu$K while the difference map has a mean of $+8.5~\mu$K. Furthermore, there are large scale features (particularly in regions of known strong Galactic emission) that can be seen in the difference map, that are likely to be due to foreground modeling errors and residual monopoles/dipoles.

\begin{figure}[!h]
\mbox{\epsfig{figure=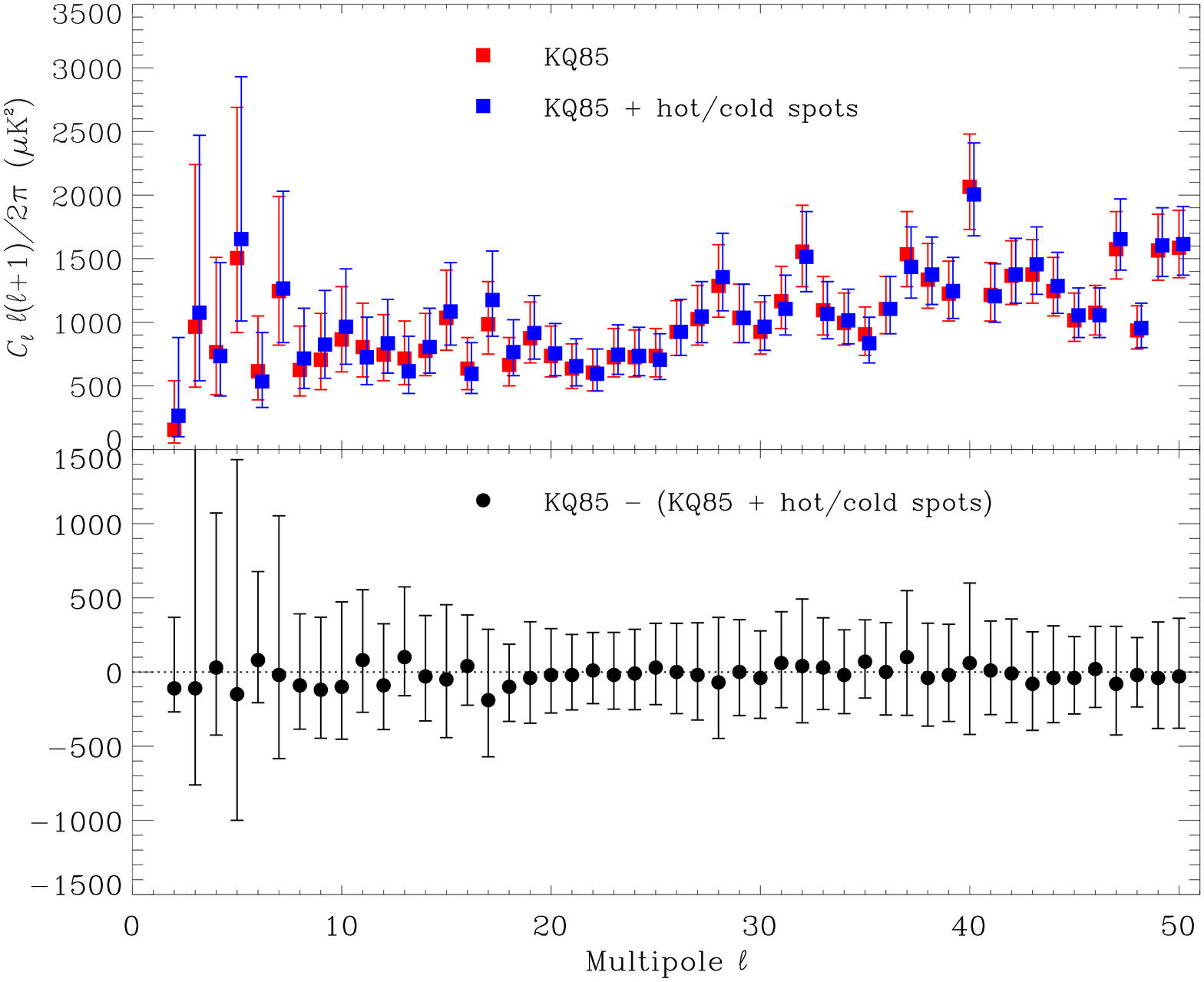,width=1.0\linewidth,clip=,angle=0}}
\caption{{\it Top}: Comparison of the CMB temperature power spectrum when using the KQ85 mask, and KQ85 mask plus the additional hot/cold spot detections of \cite{pietrobon:2008}. {\it Bottom}: Difference power spectrum.}
\label{fig:cls_kq85_kq85hotcold}
\end{figure}

Recently, \cite{pietrobon:2008} used needlets to detect non-isotropic features in the CMB sky i.e. non-Gaussianity. By masking out the brightest ($3\sigma$) detections of hot and cold spots, they claimed that a significant reduction in the CMB power occurred between $\ell=8-30$ at the $1\sigma$ level. We repeated our analysis using the additional areas masked out by the hot/cold spots. The CMB temperature power spectra are compared in  Fig.~\ref{fig:cls_kq85_kq85hotcold}. The spectra are in good agreement with the standard KQ85 mask, except at $\ell \lesssim 20$ where we confirm a slight decrease $(\approx 100~\mu$K$^2$) in power.  

\subsection{North vs South ecliptic asymmetry}
\label{sec:ecliptic}

A number of analyses of {\it WMAP} data have revealed departures from statistical isotropy on large angular scales. One of the most apparent, is the CMB power asymmetry between the northern and southern ecliptic hemispheres. Besides for possible low-l alignments, there appears to be more power in the southern hemisphere, than in the north (see for example, \cite{hansen:2004,hansen:2008,eriksen:2004,eriksen:2007,bernui:2008}). 

\begin{figure*}[]
\mbox{\epsfig{figure=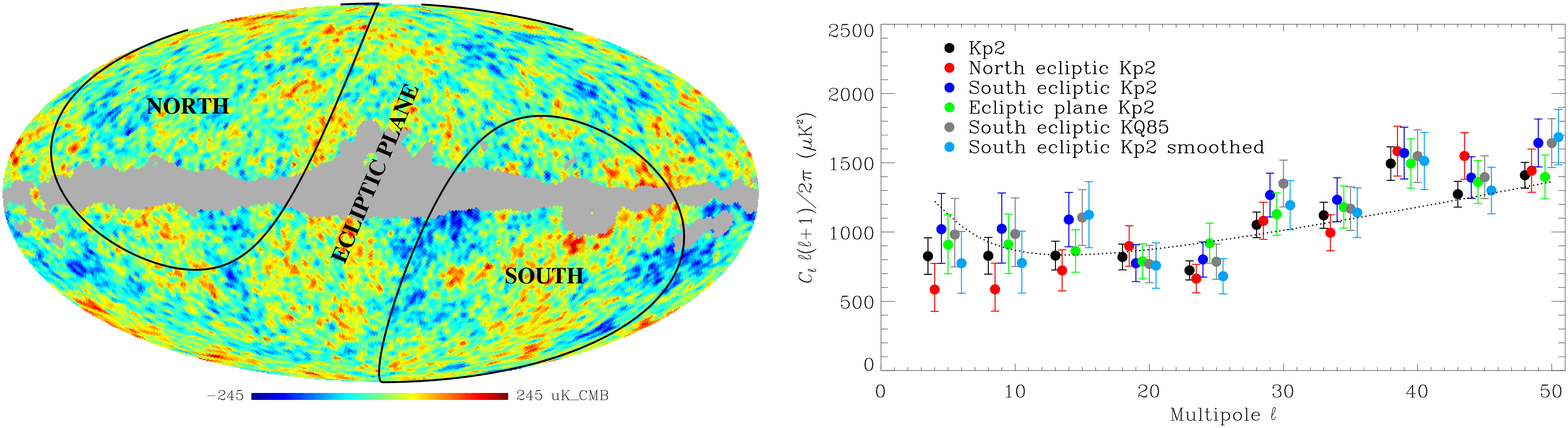,width=1.0\linewidth,clip=,angle=0}}
\caption{{\it Left}: \commander~CMB map, with the Kp2 mask region in grey. The ecliptic regions (defined by ecliptic latitudes above or below $|30^{\circ}|$) are marked. {\it Right}: \commander~power spectra for the standard Kp2 mask, and various ecliptic regions, combined with Galactic masks. Each data point is the power spectrum in $\ell$-bins with $\Delta \ell=5$. Symbols are shifted in $\ell$ for clarity.}
\label{fig:cls_ecliptic}
\end{figure*}

We ran {\tt Commander} with the same basic model as used in \S\ref{sec:wmap5_basic}, using a number of different masks for the ecliptic areas, based on a cut at ecliptic latitude $30^{\circ}$ e.g. the North ecliptic Kp2 covers ecliptic latitudes $>30^{\circ}$ in addition to the Galactic Kp2 mask, while the ecliptic plane is for latitudes within $30^{\circ}$ of the ecliptic plane. The left plot in Fig.~\ref{fig:cls_ecliptic} shows the {\tt Commander} CMB map, with Kp2 mask in grey, and ecliptic region boundaries shown as black lines. We also tried the same using the {\it WMAP} KQ85 temperature mask and the $10^{\circ}$ smoothed Kp2 mask (see \S\ref{sec:masks}).

Fig.~\ref{fig:cls_ecliptic} shows the power spectra for a selection of masks, based on the Galaxy and ecliptic latitudes. For clarity, we plot binned power spectra with $\Delta \ell=5$. The North-South asymmetry is visible, mostly at $\ell \lesssim 15$. The three lowest $\ell$-bins are $\approx 2\sigma$ discrepant between the north and south ecliptic hemispheres. The CMB in the ecliptic plane is in between, but closer to the southern values. Interestingly, the power spectra for the southern ecliptic hemisphere, when using a larger Kp2 smoothed Galactic mask contains less power at $\ell \lesssim 10$. This can be understood by simply looking at the CMB map and observing where the large-scale features lie relative to the ecliptic (left hand of Fig.~\ref{fig:cls_ecliptic}). The two bright large-scale features are both near the edge of the standard Kp2 mask; a hot spot near south of the Gum nebula region ($(l,b)\approx(260^{\circ},-20^{\circ})$) and a cold spot to the south-west of the Galactic center, at ecliptic latitude $\approx -30^{\circ}$. Both of these will be largely masked out in the smoothed Kp2 mask, thus reducing the power at low-$\ell$. However, both of these features appear to remain robust against foreground subtraction and therefore remain a mystery.

\subsection{Synchrotron / low frequency foreground spectral index maps}
\label{sec:index_maps}

One of the products of running {\tt Commander}, besides the CMB posteriors, are the posteriors for the foreground parameters. In \S\ref{sec:wmap5_basic} we presented the CMB and foreground results when applied to {\it WMAP} 5-yr data. An important product is the spectral index of the low frequency foregrounds, as fitted over the {\it WMAP} frequency range, for each pixel. This accounts for all the diffuse foreground components at low frequencies, and could include contributions from synchrotron, free-free and anomalous emission. There now exists several versions of synchrotron/low-frequency spectral index maps available in electronic form. 

\begin{figure*}[]
\mbox{\epsfig{figure=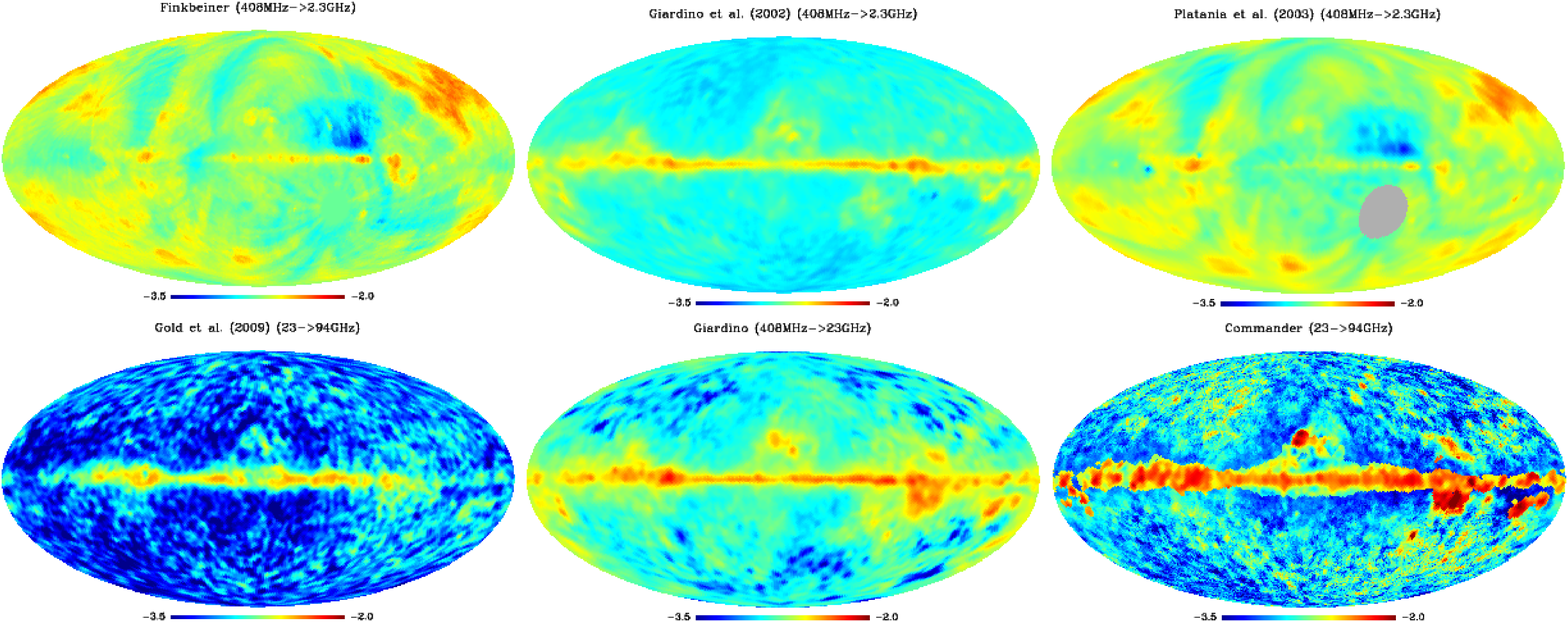,width=1.0\linewidth,clip=,angle=0}}
\caption{Comparison of synchrotron / low frequency spectral index maps. {\it Top row}: low frequency spectral index maps, from {\it left} to {\it right}: Finkbeiner 408~MHz to 2.3~GHz, Giardino et al. (2002) 408~MHz to 2.3~GHz, Platania et al. (2003) 408~MHz to 2.3~GHz. {\it Bottom row}: spectral index maps including {\it WMAP} data, from {\it left} to {\it right}: Gold et al. (2009) $23-94$~GHz, Giardino 408~MHz to 23~GHz, \commander~23-94~GHz.}
\label{fig:index_maps}
\end{figure*}

Fig.~\ref{fig:index_maps} shows spectral index maps from several authors, smoothed to a common resolution of $\sim 3^{\circ}$. Table~\ref{tab:index_maps} lists the maps, the frequency range, reference and the mean/r.m.s. spectral index at high latitude ($|b|>20^{\circ}$). The top row of maps in Fig.~\ref{fig:index_maps} were derived using low frequency data at 408~MHz \citep{haslam:1982}, 1420~MHz \citep{reich:1986} and 2326~MHz \citep{jonas:1998}. The bottom row includes {\it WMAP} data, either on their own, or combined with low frequency data (Table~\ref{tab:index_maps}).

There is considerable variation between these maps, which can be partly attributed to the different frequency ranges used, and also different assumptions that were made in their production. The mean spectral index and r.m.s. variation at high latitude ($|b|>20^{\circ}$) are given in Table~\ref{tab:index_maps}. The low frequency spectral index maps are typically flatter, with $\beta \approx -2.8$, while at {\it WMAP} frequencies, $\beta \approx -3.0$. This is in broad agreement with previous studies \citep{reich:1988,davies:1996}. \cite{deOliveira-Costa:2008} analyzed radio maps from 10~MHz to 100~GHz and found a $-2.6 < \beta < -2.3$ at 150~MHz, but $-3.3 < \beta < -2.1$ at 5~GHz. This is consistent with what is expected for synchrotron radiation, where spectral aging of cosmic ray electrons reduces the power at higher frequencies.

The low frequency spectral index maps of Finkbeiner\footnote{Software and maps available from http://astro.berkeley.edu/dust/} (priv. comm.) and \cite{platania:2003} are very similar. They used similar datasets and Fourier filtering techniques for reducing the striping artifacts that exist in the radio maps. The \cite{giardino:2002} map is somewhat different, both in absolute terms but also in morphology, due to different processing techniques, particularly for dealing with striations in the radio maps.

The \cite{gold:2009} spectral index map has a similar morphology to the \commander~map, with steeper indices towards the Galactic center region, and flatter indices outside. The \commander~map is typically flatter in the Galactic plane region, which is probably a result of using spatial templates for free-free/thermal-dust, compared to the MCMC analysis of \cite{gold:2009}. The Giardino (priv. comm.) 408~MHz-23~GHz spectral index map is somewhat different in nature. Much of this is attributed to the larger frequency range employed and the reliance on the {\it WMAP} synchrotron MEM model \citep{hinshaw:2007} as a reliable map of the synchrotron radiation\footnote{ftp://ftp.rssd.esa.int/pub/synchrotron/README.html}.

We note once again that we do not see a significant evidence for a ``{\it WMAP} Haze'', as was first seen by \cite{finkbeiner:2004a}, and later suggested to be flat spectrum synchrotron from dark matter annihilation \citep{finkbeiner:2004b,dobler:2008a}. First of all, we do not see a clear flattening of the spectral index near the Galactic center region (e.g. Fig.~\ref{fig:wmap5_maps}); in fact, the integrated foreground is typically steeper in this region outside the mask, although the ``haze'' is only a sub-component of the total emission. Secondly, there is no clear sign of an extra unmodeled component in the $\chi^2$ goodness-of-fit map of Fig.~\ref{fig:wmap5_basic_chisq}, although we do note that the $\chi^2$ does increase slightly towards the Galactic plane (as expected from the increasing complexity) and also in the ecliptic plane (due to additional unmodeled instrumental noise) which happens to pass through the Galactic center region.

\begin{table}
\centering
 \caption{Characteristics of the synchrotron / low frequency spectral index maps in Fig.~\ref{fig:index_maps}. The last column gives the mean spectral index ($\beta$), and the r.m.s. variation ($\Delta \beta$) at high latitude ($|b|>20^{\circ}$).}
\label{tab:index_maps}
  \begin{tabular}{ccc}
    \hline
Frequency        &Reference        &$\beta \pm \Delta \beta$   \\
range            &                 &$(|b|>20^{\circ})$             \\ \hline
$0.408-2.3$~GHz    &Finkbeiner (priv. comm.)      &$-2.69\pm0.12$                \\
$0.408-2.3$~GHz  &Giardino et al. (2002) &$-2.90\pm0.06$          \\
$0.408-2.3$~GHz  &Platania et al. (2003) &$-2.68\pm0.11$           \\
$23-94$~GHz      &Gold et al. (2008)  &$-3.18\pm0.17$            \\
$0.408-23$~GHz   &Giardino (priv. comm.)          &$-2.89\pm0.13$             \\
$23-94$~GHz      &This paper         &$-2.97\pm0.21$             \\
\hline      
\end{tabular}
\end{table}


\section{Conclusions} 
\label{sec:conclusions}

The Gibbs sampling algorithm provides a self-contained method to perform component separation and CMB power spectrum estimation for multi-frequency data. As implemented in the \commander~code, we can fit for a variety of signal models, including spatial templates and/or parametric functions on a per pixel basis, and propagate errors seamlessly the CMB likelihood. The method and code has now been tested and validated on a variety of simulations and real data.

In this paper, we applied the \commander~code to the 5-yr {\it WMAP} data using a signal model consisting of CMB, noise, a power-law with amplitude and spectral index in each $N_{\rm side}=64$ pixel, a spatial template for thermal dust, and monopoles/dipoles at each frequency. For the Kp2 mask, the recovered power spectrum, up to $\ell=50$, is in very good agreement with the published {\it WMAP} team spectrum and the cosmological parameters are negligibly affected. Using larger masks did not improve the power spectrum, in terms of reducing foreground residuals. However, smaller masks ($\lesssim 15\%$ masked out) did result in a bias in the power spectrum, indicating significant residuals. A full-sky analysis results in $\sim 500~\mu$K$^2$ of additional power for $\ell < 50$.

Residual monopoles and dipoles, that were originally found to be a problem with {\it WMAP} 3-yr data, are found to be small in the 5-yr data, within a few $\mu$K of zero. Although very small, not accounting for them, can affect the details of the recovered foreground parameters. The thermal dust template amplitude at 94~GHz is remarkably close (within $\approx 5$\%) to the prediction by \cite{finkbeiner:1999}, assuming a dust spectral index $\beta_{\rm dust}=+1.7$. A power-law, with amplitude and spectral index in each pixel, accounts for all the low frequency diffuse foregrounds, and is found to be an adequate model outside the Kp2 mask at {\it WMAP} frequencies. The best-fitting thermal dust spectral index, based on the $\chi^2$ goodness-of-fit statistic, appears to be much steeper than would be expected, and is likely to be an artifact of modeling errors due to dust-correlated emission with a different spectrum. Again, the effect on the CMB power spectrum is negligible, due to the low absolute levels of thermal dust at {\it WMAP} frequencies.

We tested the assumptions that were made about the signal model including the spectral index prior (mean and width), thermal dust spectral index, monopoles/dipoles. In all cases, the impact on the CMB power spectrum was very small to negligible. However, the assumptions did change the details of the recovered foreground parameters, particularly at very high latitudes, where the signal-to-noise ratio in each pixel is relatively low, and therefore the priors are having an impact. 

To account for free-free emission, we fitted for a H$\alpha$ template and found amplitudes that were lower than theory (assuming $T_{e} \approx 8000$~K), but is similar to those found by other authors. The effect on the CMB map is small ($<1~\mu$K) for most of the sky but rises to $\sim 5-10~\mu$K at lower latitudes near the edge of the Kp2 mask. Similarly, the effect on the map of low frequency spectral index is also small ($\Delta \beta \lesssim 0.1$) except near well-known star forming regions (e.g. Orion, Gum nebula) where free-free emission dominates at {\it WMAP} frequencies. This is as expected, since a single power-law cannot account for two components (e.g. synchrotron and free-free) that are approximated by power-laws with different indices.

When fitting for templates only (i.e. no spatial variation in spectral index), we find that using a combination of 408~MHz, H$\alpha$ and FDS99 templates, results in residual foreground power at low $\ell$. This indicates that this model is not sufficient to account for all the foreground complexities, particularly at the lower {\it WMAP} frequency channels. We found that the template amplitudes were sensitive to the details of the Galactic mask. The FDS99 template amplitudes rise at the lower {\it WMAP} channels, indicative of spinning dust, and similar to previous analyses. Replacing the 408~MHz template with the K$-$Ka-band difference map, and fitting to Q-, V-, W-bands, results in a cleaner separation and unbiased power spectrum.

The \commander~CMB map was compared with maps available in the literature. Differences of $\sim 10-20~\mu$K were observed, and can be attributed to the differences in techniques and assumptions of component separation. The asymmetry of power between the north and south ecliptic hemispheres is confirmed and remains robust against foreground modeling. The \commander~low frequency spectral index has a similar morphology and range to the MCMC analysis of \cite{gold:2009}, with $\beta=-2.97\pm0.21$ at high latitude. The spectral index is typically steeper by $\Delta \beta \approx 0.2$ than is observed at low ($\sim 1$~GHz) frequencies. 

\commander~will be a valuable low-$\ell$ tool for interpreting future CMB data, such as the upcoming {\it Planck} mission. The increased sensitivity, and more importantly, larger frequency range will allow a more complete foreground model to be fitted for, and should avoid the need for spatial templates \citep{leach:2008}. In a forthcoming paper, we will investigate the effectiveness of {\it WMAP}+{\it Planck} data when fitting for CMB and multiple foreground components. The increased sensitivity in polarization will allow the same procedure to be applied to the Q and U polarization maps.


\begin{acknowledgements}
  We acknowledge use of the HEALPix software \citep{gorski:2005} and
  analysis package for deriving the results in this paper. We
  acknowledge use of the Legacy Archive for Microwave Background Data
  Analysis (LAMBDA). This work was partially performed at the Jet
  Propulsion Laboratory, California Institute of Technology, under a
  contract with the National Aeronautics and Space Administration. We acknowledge the use of the NOTUR super computing facilities, the IPAC {\it Planck} cluster, and the Titan cluster owned and maintained by the University of Oslo. CD acknowledges support from the U.S. {\it Planck}
  project, which is funded by the NASA Science Mission Directorate. The work of CD was also supported in part by a STFC Advanced Fellowship. HKE
  acknowledges financial support from the Research Council of Norway. BDW was partially supported by NSF-AST 0507676 and NASA JPL 1236748.
\end{acknowledgements}

\end{document}